\documentclass[12pt]{article}

\usepackage{epsfig}
\usepackage{amsmath}
\usepackage{amsfonts}
\usepackage{graphicx}
\usepackage[german, english]{babel}
\usepackage{a4wide}
\usepackage{amsmath}
\usepackage{amssymb}
\usepackage{ifthen}
\usepackage{epsfig}

\newcommand{\beq}{\begin{equation}}
\newcommand{\eeq}{\end{equation}}
\newcommand{\beqs}{\begin{eqnarray}}
\newcommand{\eeqs}{\end{eqnarray}}

\newcommand{\nn}{\nonumber}

\newcommand{\bfR}{{\Bbb R}}

\newtheorem{oldtheorem}{Theorem}[section]
\newtheorem{oldassertion}[oldtheorem]{Assertion}
\newtheorem{oldproposition}[oldtheorem]{Proposition}

\newtheorem{oldlemma}[oldtheorem]{Lemma}
\newtheorem{olddefinition}[oldtheorem]{Definition}
\newtheorem{oldclaim}[oldtheorem]{Claim}
\newtheorem{oldcorollary}[oldtheorem]{Corollary}

\newenvironment{theorem}{\begin{oldtheorem}$\!\!\!${\bf.}}{\end{oldtheorem}}

\newenvironment{lemma}{\begin{oldlemma}$\!\!\!${\bf.}}{\end{oldlemma}}

\newbox\qedbox
\newenvironment{proof}{\smallskip\noindent{\bf Proof.}\hskip \labelsep}%
                        {\hfill\penalty10000\copy\qedbox\par\medskip}

\newcommand{\insertplot}[5]{\begin{figure}
 \hfill\hbox to 0.05in{\vbox to #5in{\vfill
 \inputplot{#1}{#4}{#5}}\hfill}
 \hfill\vspace{-.1in}
 \caption{#2}\label{#3}
 \end{figure}}
 \newcommand{\inputplot}[3]{
 \special{ps: plotfile #1}
}
\newcounter{fig}

\begin{document}
\begin{center}

{\Large Spherically symmetric selfdual Yang-Mills instantons on curved
backgrounds in all even dimensions}
\vspace{0.6cm}
\\
Eugen Radu$^{\dagger}$, D. H. Tchrakian$^{\star\dagger}$ and Yisong
Yang$^{\ddagger \, \diamond}$
\\
\vspace{0.4cm}
$^{\dagger  }${\small School of Theoretical Physics -- DIAS, 10
Burlington
Road, Dublin 4, Ireland }
\\
$^{\star}${\small Department of
Mathematical Physics, National University of Ireland, Maynooth, Ireland}
\\
$^{\ddagger}${\small  Department of Mathematics, Polytechnic
University, Brooklyn, New York 11201, USA}\\
 $^{\diamond}${\small Chern Institute of Mathematics, Nankai
 University, Tianjin 300071, PR China}
\end{center}
\newcommand{\dd}{\mbox{d}}
\newcommand{\tr}{\mbox{tr}}
\newcommand{\la}{\lambda}
\newcommand{\ka}{\kappa}
\newcommand{\al}{\alpha}
\newcommand{\ga}{\gamma}
\newcommand{\de}{\delta}
\newcommand{\si}{\sigma}
\newcommand{\bomega}{\mbox{\boldmath $\omega$}}
\newcommand{\bsi}{\mbox{\boldmath $\sigma$}}
\newcommand{\bchi}{\mbox{\boldmath $\chi$}}
\newcommand{\bal}{\mbox{\boldmath $\alpha$}}
\newcommand{\bpsi}{\mbox{\boldmath $\psi$}}
\newcommand{\brho}{\mbox{\boldmath $\varrho$}}
\newcommand{\beps}{\mbox{\boldmath $\varepsilon$}}
\newcommand{\bxi}{\mbox{\boldmath $\xi$}}
\newcommand{\bbeta}{\mbox{\boldmath $\beta$}}
\newcommand{\ee}{\end{equation}}
\newcommand{\eea}{\end{eqnarray}}
\newcommand{\be}{\begin{equation}}
\newcommand{\bea}{\begin{eqnarray}}
\newcommand{\ii}{\mbox{i}}
\newcommand{\e}{\mbox{e}}
\newcommand{\pa}{\partial}
\newcommand{\Om}{\Omega}
\newcommand{\vep}{\varepsilon}
\newcommand{\bfph}{{\bf \phi}}
\newcommand{\lm}{\lambda}
\def\theequation{\thesection.\arabic{equation}}
\renewcommand{\thefootnote}{\fnsymbol{footnote}}
\newcommand{\re}[1]{(\ref{#1})}
\newcommand{\R}{{\rm I \hspace{-0.52ex} R}}
\newcommand{\N}{{\sf N\hspace*{-1.0ex}\rule{0.15ex}%
{1.3ex}\hspace*{1.0ex}}}
\newcommand{\Q}{{\sf Q\hspace*{-1.1ex}\rule{0.15ex}%
{1.5ex}\hspace*{1.1ex}}}
\newcommand{\C}{{\sf C\hspace*{-0.9ex}\rule{0.15ex}%
{1.3ex}\hspace*{0.9ex}}}
\newcommand{\Pee}{{\sf P\hspace*{-0.9ex}\rule{0.15ex}%
{1.3ex}\hspace*{0.9ex}}}
\newcommand{\eins}{1\hspace{-0.56ex}{\rm I}}
\renewcommand{\thefootnote}{\arabic{footnote}}

%\maketitle

\begin{abstract}
We present several different classes of selfdual Yang-Mills
instantons in all even $d$ backgrounds with
Euclidean signature. In $d=4p+2$ the only solutions we
found are on constant curvature dS and AdS backgrounds, and are evaluated in
closed form. In $d=4p$ an interesting class of instantons are
given on black hole backgrounds. One
class of solutions are (Euclidean) time-independent and
spherically symmetric in $d-1$ dimensions, and the other class are
spherically symmetric in all $d$ dimensions. Some of the solutions
in the former class are evaluated numerically, all the rest being
given in closed form. Analytic proofs of existence covering all
numerically evaluated solutions are given. All instantons studied
have finite action and vanishing energy momentum tensor and do not
disturb the geometry.
\end{abstract}

\medskip
%%%%%%%%%%%%%%%%%%%%%%%%%%%%%%%%%%%%%%%%%%%%%%
\section{Introduction}
\setcounter{equation}{0}
%%%%%%%%%%%%%%%%%%%%%%%%%%%%%%%%%%%%%%%%%%%%%
The study of selfdual solutions of Yang-Mills (YM) theory on curved backgrounds
has proven to be a fruitful field of research in physics and mathematics.

While most recent work on gravitating YM theory has been carried out
in Lorentzian signature spacetimes, the earliest work on the
subject, carried out by Charap and Duff~\cite{Charap:1977ww},
Chakrabarti and collaborators~\cite{Chakrabarti:1987kz}, was in four
dimensional ($d=4$) spacetimes of Euclidean signature. This was
quite natural, as a sequel to the study of gravitational
instantons~\cite{Eguchi:1980jx}. In both \cite{Charap:1977ww} and
\cite{Chakrabarti:1987kz}, the YM connection $A_{\mu}$ is identified
with the (gravitational) spin--connection $\omega_{\mu}^{mn}$ as \be
\label{emb}
A_{\mu}=-\frac12\,\omega_{\mu}^{mn}\,\Sigma_{mn}^{(\pm)}\quad\Rightarrow\quad
F_{\mu\nu}=-\frac12\,R_{\mu\nu}^{mn}\,\Sigma_{mn}^{(\pm)}\,, \ee
$F_{\mu\nu}$ and $R_{\mu\nu}^{mn}$ being the YM and the Riemann
curvatures, respectively, and $\Sigma_{mn}^{(\pm)}$ one or other of
the chiral representations of the algebra of $SO(4)$, {\it i.e.}
left or right $SU(2)$. In both
cases~\cite{Charap:1977ww,Chakrabarti:1987kz}, the instantons are
{\it selfdual} in the YM curvature, and are evaluated in closed
form. Selfduality of the YM curvature results in the vanishing of
the stress tensor as a function of the non-Abelian matter fields, so
that the latter has no backreaction on gravity, {\it i.e.} these
instantons are essentially given on a fixed curved background.

What is special about the $d=4$ Charap--Duff (CD) instanton in
\cite{Charap:1977ww}, is that the Riemann curvature is also {\it
double-selfdual}, which fixes the form metric of the metric
background ($e.g.$ the Euclideanised Schwarzschild background for
the solution in \cite{Charap:1977ww}). However, we argue in this
work that instanton configurations with rather similar properties
exist for any  spherically symmetric metric satisfying a suitable
set of boundary conditions (this includes $e.g.$ the
Reissner-Nordst\"om background). Although a closed form solution is
found for a Schwarzschild metric only, we present existence proofs
for the solutions we found numerically. In the present work, we will
refer to this type of instantons (and it generalizations) as
solutions of Type I.

Further to these (Euclidean time) static instantons
\cite{Charap:1977ww,{Chakrabarti:1987kz}}, a new type of $d=4$
static YM instanton on a curved background was recently discovered
in \cite{Brihaye:2006bk} to which we shall refer as Type II instantons.
These are basically deformed Prasad--Sommerfield~\cite{Prasad:1975kr} (PS) monopoles.
Like the Type I instantons, the solutions in~\cite{Brihaye:2006bk}
are also selfdual, but differ in an essential way from Type I
instantons, in that they satisfy different boundary conditions
and have a different action for the same background.
As conjectured in \cite{Brihaye:2006bk}, the Type II instantons
exist for an arbitrary nonextremal $SO(3)$-spherically symmetric background,
the PS solution
being recovered in the $R^3\times S^1$ flat space limit.
The actions of both Types I and II instantons saturate
the bound of the usual $2$nd Chern--Pontryagin (CP) charge.
They are both given on fixed Euclideanised black hole backgrounds.

The larger part of this paper is concerned with the generalization
of the $d=4$ solutions of both Types I and II to arbitrary even
dimensions\footnote{Restriction to even dimensions is because of our
requirement of {\it selfduality}, without including Higgs or other
scalar matter fields.}. These instantons are static and spherically
symmetric in $d-1$ dimensions and have a vanishing stress tensor. We
argue that the form metric backgrounds are not crucial for the
existence of  these solutions, as long as the metric functions
satisfy a rather weak set of conditions. Here we will consider
mainly Schwarzschild like backgrounds with and without a
cosmological constant, and with a $U(1)$ field, presenting also an
existence proof for a more general case.

In addition to these static solutions, we also study YM instantons which
are spherically symmetric in the full $d$ dimensional Euclidean spacetime.
These are deformations of the BPST instanton~\cite{BPST}, and are likewise
selfdual, and hence are also solutions on a fixed curved background. In the case of AdS$_{4}$
background\footnote{Selfdual instantons on compact symmetric backgrounds,
as opposed to ones on the noncompact space AdS$_{4}$, were known.
For example YM instantons on $S^4$ were constructed by Jackiw and
Rebbi~\cite{Jackiw:1976dw}, and those on $\C\Pee^2$,
by Gibbons and Pope~\cite{Gibbons:1978zy} a long
time ago.} this was given recently by Maldacena and Maoz~\cite{Maldacena:2004rf}
\footnote{In \cite{Maldacena:2004rf} also wormhole
solutions to the second order equations, where the matter field curves the geometry,
are given, which in the dS$_{4}$ case were already known~\cite{Hosoya:1989zn}.
Here we have restricted to selfdual instantons.}.
The deformed BPST instantons on AdS$_{4}$ and dS$_{4}$ are
generalised to AdS$_{d}$ and dS$_{d}$ for all
even $d$, the new solutions~\footnote{These are the instantons on noncompact
symmetric spaces, corresponding to
the already known ones on the compact spaces, namely on $S^{4p}$ in \cite{O'Se:1987fx},
on $S^{2n}$ in \cite{O'Brien:1988rs,Kihara:2007di}, and on on $\C\Pee^n$ in \cite{Ma:1990ja}.}
in $d=4p$ being deformations of the BPST hierarchy \cite{Tchrakian:1984gq}.

In general, gravitating YM instantons in higher dimensions are of physical
relevance in the study of field theories arising from superstring
theory~\cite{Green:1987sp,Polchinski:1998rq}.
In particular, a special aspect
of selfdual instantons is that they can be employed in supersymmetric
gravity theories, for example in the
analysis of branes in $4p+1$ dimensions generalising that of $5$-Branes proposed in
Ref.~\cite{Strominger:1990et}
(see also Gibbons {\it et. al.}~\cite{Gibbons:1993xt}).
More recently, {\it six} dimensional instantons were employed by Kihara and Nitta~\cite{Kihara:2007vz}
for Cremmer--Scherck compactification over $S^6$, which can also be
generalised to all {\it even} dimensions.

%In the present work we will be mainly concerned with {\it selfdual}~\footnote{In fact we searched for {\it nonselfdual} radial excitations of the {\it selfdual} instantons presented here, numerically, but failed to find any.} YM instantons in arbitrary even for the YM fields, without introducing Higgs or other scalar matter fields.} spacetime dimensions with Euclidean signature.

Instantons of non-Abelian field systems in dimensions higher than
$4$ can be constructed for the hierarchy~\cite{Tchrakian:1992ts} of
Yang--Mills models in all even dimensions, \be \label{YMhier} {\cal
L}_{\rm{YM}}^{(P)}=\sum_{p=1}^{P}\frac{\tau_p^2}{2(2p)!}\
\mbox{Tr}\,F(2p)^2\;, \ee in which the $2p$-form $F(2p)$ is the
$p$-fold antisymmetrised product $F(2p)=F\wedge F\wedge...\wedge F$
of the YM curvature $2$-form $F$, and we choose the YM connections
to take their values in the chiral representations $SO_{\pm}(d)$ of
$SO(d)$. Here the maximum value of $P$ in the superposition
\re{YMhier} is simply $P_{\rm{max}}\le\frac12d.$ Such instantons are
not necessarily selfdual. In particular, when all but the
$p=\frac14d$ term in \re{YMhier} is retained, one finds a hierarchy
of BPST instantons~\cite{Tchrakian:1984gq} as well as the Witten
type solutions \cite{1,2,3} in $d=4p$ dimensions, which satisfy the
corresponding selfduality equation \be \label{sd4p}
F(2p)=\pm\,^{\star}F(2p)\,, \ee $^{\star}F(2p)$ being the Hodge dual
of $F(2p)$, including the appropriate factor of $e=\sqrt{{\rm
det}\,g}$.

The selfduality equations \re{sd4p} feature higher orders of the YM $2-$form curvature.
If one restricts to the usual ($p=1$) YM model in higher dimensions, the action will
be infinite. There are several other selfduality equations defined on higher {\it even}
dimensions in the literature, which are linear in the YM curvature. But for none of these does
the (usual) YM action saturate a topological lower
bound and result in infinitely large action. Such selfduality equations are irrelevant for
our purposes here. Also, the solution to the $p=2$ member of \re{sd4p}
results in infinite action if it is not recognised~\cite{Grossman:1984pi} that
the Lagrangian is the
$p=2$ member of the YM hierarchy in \re{YMhier}, and not the usual $p=1$ member.
It can be noted that
there is a another hierarchy of, nonlinear in the YM curvature, selfduality
equations~\cite{Saclioglu:1986qn} defined in all even dimensions.
In $4p$ dimensions, this hierarchy
coincides with the \re{sd4p} of \cite{Tchrakian:1984gq}, saturating
the action of the $p=\frac14d$ in
\re{YMhier}. In $4p+2$ dimensions however, the Lagrangians of
\cite{Saclioglu:1986qn,Fujii:1986ty,Fujii:1986tz}, whose field equations
these nonlinear selfduality
equations~\cite{Saclioglu:1986qn} solve, come in odd powers
of the YM curvature and hence are not bounded from below. We therefore
 restrict our attention to the
hierarchy \re{YMhier} henceforth, both in $4p$ and $4p+2$ dimensions.

In fact, since we are in effect concerned only with selfdual solutions,
we will only ever consider two special cases of the YM hierarchy \re{YMhier}.
In $d=4p$ this is the system
consisting of a single term with $p=\frac14d$ saturated by \re{sd4p}.
In $d=2(p+q)$, with $q\neq p$,
there are two terms in \re{YMhier}
labeled by $p$ and $q$. The system in this case is saturated by the selfduality equations
\be
\label{sdpq}
\tau_p\,F(2p)=\pm\,\tau_q\,(^{\star}F(2q))(2p)\,,
\ee
where the Hodge dual of the $2q$ form on the right hand side of \re{sdpq} is a $2p$ form,
with an obvious
relation between the dimensions of the constants $\tau_p$ and $\tau_q$.

The choice of the hierarchy \re{YMhier} consisting of higher order terms in the YM curvature
can be justified in the light of the presence of such terms in the low energy string theory
(see e.g. \cite{Tseytlin}-\cite{CNT}) Lagrangian.

The imposition of spherical symmetry in $d-1$ spacelike dimensions
for (Euclidean) time static fields is presented in section {\bf 2},
while the more compact task of imposing spherical symmetry in all
$d$ Euclidean dimensions is deferred to section {\bf 5} where such
instantons are constructed. The static solutions which are
spherically symmetric in the $d-1$ dimensional subspace are
presented in section {\bf 3}, while those in section {\bf 5} are
spherically symmetric in all $d$ dimensions. All instantons
presented in section {\bf 5} are given in closed form. Section {\bf
3} is divided into two parts. In the first subsection, {\bf 3.1},
Type I {\it selfdual} instantons (generalising the $d=4$ CD
instanton~\cite{Charap:1977ww}) are evaluated in closed form for
double-selfdual gravitational backgrounds. Also in subsection {\bf
3.1}, solutions satisfying Type I boundary conditions, but {\it not}
on double-selfdual gravitational backgrounds, are constructed
numerically. In subsection {\bf 3.2} Type II solutions, which
satisfy boundary conditions that differ from those of Type I
solutions, are presented. These are evaluated exclusively
numerically. To underpin the numerically constructed solutions,
analytic proofs for their existence are given in section {\bf 4}.
%Finally in section {\bf 5}, we present closed form solutions
%that are spherically symmetric in the full $d$ dimensions, both for $d=4p$ and $d=4p+2$,
All the solutions presented are selfdual satisfying the hierarchy of selfduality equations
\re{sd4p} and \re{sdpq}, respectively. In the case of Types I and II instantons
the second order equations were integrated numerically in search of radial excitations,
and none were found. A summary and discussion of our results is given in section {\bf 6}.
Finally, an analysis of double--selfdual spaces
is given in the Appendix, since these play an important role in the
construction of the Charap--Duff hierarchy.

%%%%%%%%%%%%%%%%%%%%%%%%%%%%%%%%%%%%%%%%%%%%%%
%\section{Symmetry imposition}
%\setcounter{equation}{0}
\section{Symmetry imposition: spherical symmetry in $d-1$ dimensions}
\setcounter{equation}{0}
%%%%%%%%%%%%%%%%%%%%%%%%%%%%%%%%%%%%%%%%%%%%%

 In this section, we impose spherical
symmetry in $d-1$ dimensional subspace on the (Euclidean time)
static gravitational and gauge fields.

%%%%%%%%%%%%%%%%%%%%%%%%%%%%%%%%%%%%%%%%%%%%%%
 \subsection{General results}
%%%%%%%%%%%%%%%%%%%%%%%%%%%%%%%%%%%%%%%%%%%%%
We consider a metric Ansatz with spherical symmetry in $d-1$ dimensional subspace,
%The spherically symmetric metric Ansatz is parameterised by two functions $N(r)$ and $\sigma(r)$
\be
\label{metric}
ds^2 = N(r)\sigma^2(r) d\tau^2\ +\ N(r)^{-1} dr^2\
+\ r^2 d \Omega_{(d-2)}^2 \: .
\ee
Here $d\Omega_{(d-2)}^2$ is the metric on a $(d-2)$-dimensional sphere, $\tau$ corresponds to the Euclidean
time, while $r$ is the radial coordinate. We shall be mainly interested in asymptotically flat background
metrics whose fixed point set of the Euclidean time symmetry  is of $d-2$ dimensions (a "bolt") and
the range of the radial coordinate is restricted to $r_h\leq r<\infty$, while
\begin{eqnarray}
\label{m-eh}
%\nonumber
N(r)=N_1(r-r_h)+N_2(r-r_h)^2+O(r-r_h)^3,
~~~
\sigma(r)=\sigma_h+ \sigma_1(r-r_h)+O(r-r_h)^2,
\end{eqnarray}
where $N_1,~N_2~\sigma_h,~\sigma_2$ are  constants
 determined
by the equations of motion\footnote{
For most of this section we are not interested in the precise
form of the functions $N$ and $\sigma$, the
considered YM instantons presenting some generic features
for any choice of the background compatible with this behaviour.}
(with $N_1,~\sigma_h$ positive quantities).

This type of metric usually corresponds to the analytical
continuations of Lorentzian black hole solutions.
The absence of conical singularities at $r=r_h$ fixes the periodicity of the coordinate $\tau$
\begin{eqnarray}
\label{beta}
\beta=\frac{4\pi}{\sigma_h N_1}~.
\end{eqnarray}
(Note that this holds for any gravity-matter model we consider.)
As $r\to \infty$, the Euclideanised (thermal-)Minkowski background
is approached, with  $\sigma(r) \to 1$, $N(r) \to 1-(r_0/r)^{k}$,
with $r_0$ a positive constant, the value of $k$ depending on the
gravity model we are using (e.g. $k=d-3$ for the usual Einstein gravity).

Since some of the numerical work in section {\bf 3} is carried out for $p-$Einstein backgrounds
defined for the system \re{EHhier} with $p=q$, we state the reduced one dimensional gravitational Lagrangian
in $d$ spacetime subject to the static spherically symmetric metric Ansatz~\re{metric}
\be
\label{redlaggrav}
L^{(p,d)}_{(\rm{grav})}=
\frac{\ka_p}{2^{2p-1}}\,\frac{(d-2)!}{(d-2p-1)!}\ \si\,
\frac{d}{dr}\left[r^{d-2p-1}(1-N)^p\right]\,.
\ee

Next, we impose spherical symmetry in $d-1$ dimensions on the static
YM connection $A_{\mu}=(A_0,A_i)$, $i=1,2,...,d-1$ and $\mu=0,i$, resulting in the following Ansatz
\begin{equation}
\label{YMsph}
A_0=u(r)\,\hat x_j\,\Sigma_{j,d}^{(\pm)}\ ,\quad
A_i=\left(\frac{1-w(r)}{r}\right)\,\Sigma_{ij}^{(\pm)}\hat x_j\ , \quad
\Sigma_{ij}^{(\pm)}=
-\frac{1}{4}\left(\frac{1\pm\Gamma_{d+1}}{2}\right)
[\Gamma_i ,\Gamma_j]\ ,
\end{equation}
described by two
functions $w(r)$ and $u(r)$ which we shall refer to as magnetic and electric potential, respectively.
The $\Gamma$'s denote the $d$-dimensional gamma matrices, and $\Gamma_{d+1}$, the
chiral matrix in that dimension. The radial variable in \re{YMhier} is $r=\sqrt{|x_i|^2}$ and
$\hat x_i=x_i/r$ is the unit radius vector, while $x_0=t$.

Inserting the YM Ansatz \re{YMsph} in the $p$-th term in \re{YMhier}, we have the corresponding
term in the resulting reduced one dimensional YM Lagrangian
\bea
L^{(p,d)}_{\rm{YM}}&=&\frac{\tau_p}{2\cdot (2p)!}
\frac{(d-2)!}{(d-2p-1)!}\,r^{d-4p}\Bigg\{\,\si\,(1-w^2)^{2(p-1)}
\left[(2p)N\,w'^2+\frac{(d-2p-1)}{r^2}\,(1-w^2)^2\right]
\nonumber
\\
&+&\frac{2p}{2p-1}\,\frac{1}{\si}
\left[\frac{\left(\left[(1-w^2)^{p-1}u\right]'\right)^2}{d-2p}\,r^2+
\frac{2p-1}{N}\,\left[(1-w^2)^{p-1}u\right]^2\,w^2\right]\Bigg\},
\label{LYMpd}
\eea
where a prime denotes the derivative with respect to $r$.

We now adapt the expression \re{LYMpd} to the relevant
models in $d=4p$ and $d=2(p+q)$, ($p\neq q$).
%%%%%%%%%%%%%%%%%%%%%%%%%%%%%%%%%%%%%%%%%%%%%%%%%%%%%%%%%%%%%%%%%%%%%%%%%%%%%%%%%%%%%%%%%%%%%%%%%%%%%%
\subsection{$d=4p$}
%%%%%%%%%%%%%%%%%%%%%%%%%%%%%%%%%%%%%%%%%%%%%%%%%%%%%%%%%%%%%%%%%%%%%%%%%%%%%%%%%%%%%%%%%%%%%%%%%%%%%%
In this case the action density corresponding to \re{YMhier}
with only one term $p=\frac14d$ can be
written as a sum of complete squares plus (or minus) a total derivative
\bea
L^{(p,d=4p)}_{\rm{YM}}&=&\frac{\tau_p}{2\cdot (2p)!}
\frac{(4p-2)!}{(2p-1)!}\Bigg\{
\left[\frac{r}{\sqrt{\si(2p-1)}}\,\left[(1-w^2)^{p-1}\,u\right]'\mp
\frac{\sqrt{\si(2p-1)}}{r}\,(1-w^2)^p\right]^2
\nonumber
\\
&+&2p\,(1-w^2)^{2(p-1)}\left[\sqrt{\si N}\,w'\pm\frac{1}{\sqrt{\si N}}\,u\,w\right]^2\pm
\frac{d}{dr}\left[(1-w^2)^{2p-1}\,u\right]\Bigg\},
\label{quad}
\eea
implying that the second order YM equations are solved by the following (anti-)selfduality equations
\bea
\sqrt{\si N}\,w'&\pm&\frac{1}{\sqrt{\si N}}\,u\,w=0
\label{4psd1}
\\
\frac{r}{\sqrt{\si(2p-1)}}\,\left[(1-w^2)^{p-1}\,u\right]'&\mp&
\frac{\sqrt{\si(2p-1)}}{r} \,(1-w^2)^p=0\,,
\label{4psd2}
\eea
which arise directly from the imposition of spherical symmetry \re{YMsph}-\re{metric} on the
(anti)selfduality equations \re{sd4p}.

Without any loss of generality, we will solve the selfduality equations
by taking the upper sign in
the above relations; the anti-instanton solutions are found by
reversing the sign of the electric potential.

The action of the selfdual solutions is
\begin{eqnarray}
\label{action-p}
S=\beta V_{d-2}\int_{r_h}^{\infty}dr~L^{(p,d)}_{\rm{YM}}
=
\frac{\tau_p}{2\cdot (2p)!}
\frac{(4p-2)!}{(2p-1)!}
\bigg(
 (1-w^2)^{2p-1}\,u\big|_{r=\infty} -
 (1-w^2)^{2p-1}\,u\big|_{r=r_h}
\bigg)~,
\end{eqnarray}
where $V_{d-2}$ is the area of the unit $S^{d-2}$ sphere.

We consider  in Section 3 two different sets of boundary conditions
for the first order equations (\ref{4psd1}), (\ref{4psd2}),
leading to different types of solutions and different values of the
action (\ref{action-p}).
%%%%%%%%%%%%%%%%%%%%%%%%%%%%%%%%%%%%%%%%%%%%%%%%%%%%%%%%%%%%%%%%%%%%%%%%%%%%%%%%%%%%%%%%%%%%%%%%%%%%%%
\subsection{$d=2(p+q)$,\ \ \ $p\neq q$}
%%%%%%%%%%%%%%%%%%%%%%%%%%%%%%%%%%%%%%%%%%%%%%%%%%%%%%%%%%%%%%%%%%%%%%%%%%%%%%%%%%%%%%%%%%%%%%%%%%%%%%
The YM reduced one dimensional Lagrangian in
this case is
\be
\label{lagpq}
L_{\rm{YM}}=L^{(p,d=2(p+q))}_{\rm{YM}}+L^{(q,d=2(p+q))}_{\rm{YM}}\,,
\ee
each of the two terms in which is readily read off \re{LYMpd},
with coupling strengths $\tau_p^2$ and
$\tau_q^2$ respectively. Just like \re{LYMpd} in $d=4p$ was
rewritten in the form \re{quad}, so can \re{lagpq} be cast into
the following useful form, consisting of sums of complete squares,
plus (or minus) a total derivative.
\bea
L_{\rm{YM}}&=&
\left(\tau_p\sqrt{\frac{2p}{(2q-1)!}}r^{q-p}
\sqrt{\si N}(1-w^2)^{p-1}w'\pm
\tau_q\sqrt{\frac{2q}{(2p-1)!}}r^{p-q}
\frac{1}{\sqrt{\si N}}(1-w^2)^{q-1}wu\right)^2
\nonumber
\\
&+&\left(\tau_q\sqrt{\frac{2q}{(2p-1)!}}r^{p-q}\sqrt{\si N}
(1-w^2)^{q-1}w'\pm
\tau_p\sqrt{\frac{2p}{(2q-1)!}}r^{q-p}\frac{1}{\sqrt{\si N}}
(1-w^2)^{p-1}wu\right)^2
\nonumber
\\
&+&\left(\tau_p\sqrt{\frac{2p}{(2p-1)(2q)!}}r^{q-p+1}\frac{1}
{\sqrt{\si}}\left[(1-w^2)^{p-1}u\right]'\mp
\frac{\tau_q}{\sqrt{(2p-2)!}}r^{p-q-1}\sqrt{\si}(1-w^2)^q\right)^2
\nonumber
\\
&+&\left(\tau_q\sqrt{\frac{2q}{(2q-1)(2p)!}}r^{p-q+1}\frac{1}
{\sqrt{\si}}\left[(1-w^2)^{q-1}u\right]'\mp
\frac{\tau_p}{\sqrt{(2q-2)!}}r^{q-p-1}\sqrt{\si}(1-w^2)^p\right)^2
\nonumber
\\
&&\qquad\qquad\qquad\qquad
\pm\tau_p\tau_q\frac{4(p+q)}{\sqrt{(2p)!(2q)!}}\,
\frac{d}{dr}\left[(1-w^2)^{p+q-1}\,u\right]\,.
\label{quadpq}
\eea
\re{quadpq} implies that the action of \re{lagpq} is absolutely
minimised by a set of (anti)selfduality
equations. These can be expressed most simply by redefining
the coupling strengths $\tau_p$ and $\tau_q$
in \re{lagpq} and \re{quadpq} according to
\[
\hat\tau_p=\tau_p\sqrt{(2p)!}\quad,\quad\hat\tau_q=\tau_q\sqrt{(2q)!}\,,
\]
resulting in
\bea
\hat\tau_p\,r^{q-p}\sqrt{\si N}\,(1-w^2)^{p-1}\,w'&=\mp&
\hat\tau_q\,r^{p-q}\frac{1}{\sqrt{\si N}}\,(1-w^2)^{q-1}\,wu
\label{pq1a}
\\
\hat\tau_q\,r^{p-q}\sqrt{\si N}(1-w^2)^{q-1}\,w'&=\mp&
\hat\tau_p\,r^{q-p}\frac{1}{\sqrt{\si N}}\,(1-w^2)^{p-1}\,wu
\label{pq1b}
\\
\hat\tau_p\,r^{q-p+1}\frac{1}{(2p-1)\sqrt{\si}}\,\left[(1-w^2)^{p-1}u\right]'&=\pm&
\hat\tau_q\,r^{p-q-1}\sqrt{\si}\,(1-w^2)^q
\label{pq2a}
\\
\hat\tau_q\,r^{p-q+1}\frac{1}{(2q-1)\sqrt{\si}}\,\left[(1-w^2)^{q-1}u\right]'&=\pm&
\hat\tau_p\,r^{q-p-1}\sqrt{\si}\,(1-w^2)^p\,,
\label{pq2b}
\eea
which also follow by directly imposing spherical symmetry \re{YMsph}-\re{metric} on the
selfduality equation \re{sdpq} Setting $p=q$, \re{pq1a}-\re{pq2b} and \re{sdpq} revert to
\re{4psd1}-\re{4psd2} and \re{sd4p} respectively.

%%%%%%%%%%%%%%%%%%%%%%%%%%%%%%%%%%%%%%%%%%%%%%
\section{Solutions with spherical symmetry in $d-1$ dimensions}
\setcounter{equation}{0}
%%%%%%%%%%%%%%%%%%%%%%%%%%%%%%%%%%%%%%%%%%%%%
Here we will construct the Types I and II solutions in the following two subsections, respectively.
Both these describe selfdual YM on black hole backgrounds, and differ from each other in the different
boundary conditions they satisfy respectively.

%%%%%%%%%%%%%%%%%%%%%%%%%%%%%%%%%%%%%%%%%%%%%%%%%%%%%%%%%%%%%%%%%%%%%%%%%%%
\subsection{Type I solutions: Extended Charap-Duff configurations and their deformations}
%%%%%%%%%%%%%%%%%%%%%%%%%%%%%%%%%%%%%%%%%%%%%%%%%%%%%%%%%%%%%%%%%%%%%%%%%%%
This subsection is divided in three parts, the first two pertaining to solutions in $d=4p$ and the
third in $d=2(p+q)$. In the first subsection we present closed form instantons on double--selfdual
backgrounds in $d=4p$, generalising the usual Schwarzschild black hole $d=4$, to which we refer
as $p-$Schwarzschild metrics. (These are not to be
confused with the Schwarzschild-Tangherlini metrics in higher dimensions, which are {\it not}
double--selfdual.) In the second subsection we construct numerical solutions on generic $4p$
dimensional backgrounds, which are not double--selfdual. The third subsection
is concerned with solutions in $d=2(p+q)$, which
are given in fixed symmetric spaces only, and not on black holes.

%%%%%%%%%%%%%%%%%%%%%%%%%%%%%%%%%%%%%%%%%%%%%%
\subsubsection{Type I instantons in $d=4p$ on double-selfdual backgrounds}
%%%%%%%%%%%%%%%%%%%%%%%%%%%%%%%%%%%%%%%%%%%%%%%%%

For $p=1$, the YM selfduality equations (\ref{4psd1}), (\ref{4psd2})
present a well known closed form solution, found a long time
ago by  Charap and Duff~\cite{Charap:1977ww} (CD).
This solution has been constructed for the case of
double-selfdual $p-$Schwarzschild background~\footnote{
It is worth noting that for the $p=1$ case only, a
generalisation of the CD instanton is obtained
by replacing the Schwarzschild background employed in {\bf 3.1.1} above, by the Euclideanised
Kerr background. Unfortunately, this more general solution cannot be extended to $d>4$
since no higher dimensional counterparts of the Kerr solution are known in $p-$Einstein gravity
for $p\ge 2$.}.

The generalisation of the CD solution to $d=4p$ case is given
formally in \cite{O'Brien:1988rs}, and here
we construct these solutions concretely. This is straightforward and is effected by
the replacement of the usual ($p=1$) Schwarzschild background with
the solution to the double-selfduality equation
\re{dsd} corresponding to the $p-$Einstein gravity defined by \re{EHp}.
It should be emphasised here that, using the
gravitational background of any other member of the gravitational
hierarchy other than the $p-$Einstein
gravity does not support a CD instanton solution. The YM instanton is found
by embedding the gauge connection into the gravity spin connection according to \re{emb}.
The resulting solution reads
\begin{eqnarray}
\label{ex-sol}
w(r)=-\sqrt{N(r)},~~u(r)=-\frac{1}{2}N'(r),
~~~{\rm with}~~~\sigma(r)=1\,,
\end{eqnarray}
where $N(r)$ is the metric function pertaining to the
solution of the $p-$Einstein
equations with cosmological term in $d=4p$ dimensions.
This is the $d=4p$ special case of the solution
given in \cite{Chakrabarti:2001di}, and
can be found by solving
\begin{eqnarray}
\label{N-sol}
\left(\frac{1-N(r)}{r^2}\right)^p=c_1+\left(\frac{r_0}{r}\right)^{d-1}\,,
%N(r)=1-\left((\frac{r_0}{r})^{2p-1}+c_1r^{2p}\right)^{1/p},~~\sigma(r)=1.
\end{eqnarray}
which result from substitution of the metric Ansatz \re{metric} in the
double-selfduality equation \re{dsd}. $r_0$ here is
related to the mass of the solution, $c_1$ being fixed by the cosmological constant.
% This is what is expected~\cite{O'Brien:1988rs} when the
%embedding \re{emb} is employed in the double-selfduality equation \re{dsd},
%whose metric satisfies the
%corresponding $p-$Einstein equation \re{peinstein} resulting from the $p-$Einstein
%Lagrangian \re{EHp}
%incorporating a cosmological constant $\Lambda$.
This result, namely that the double-selfdual metric with
Euclidean signature~\footnote{That for Minkowskian signature double--selfduality of the metric
does not lead to the $p-$Einstein equation is seen from \re{constrp} and \re{weakp} of the
Appendix.} supports a YM instanton in the presence of a cosmological constant, agrees with that
of Julia {\it et. al.}~\cite{Julia:2005ze} in $d=4$.

One can see that the gauge potentials diverge for solutions with AdS
asymptotics which leads to a diverging action, according to
(\ref{action-p}.)
For a vanishing cosmological constant
$c_1=0$, these solutions have a finite action
\begin{eqnarray}
\label{action-II}
S= \frac{\tau_p}{2\cdot (2p)!}
\frac{(4p-2)!}{(2p-1)!}  2 \pi V_{d-2}~.
\end{eqnarray}
(One can see that the background features do no enter here).
Another interesting case is provided by dS instantons.
Here the radial coordinate has a finite range and in the general case
the spacetime presents a conical singularity at $r=r_h$ or $r=r_c$
(with $r_h<r_c$, $N(r_h)=N(r_c)=0$).
The action of these solutions is
\begin{eqnarray}
\label{action-II-dS}
S=
\frac{\tau_p}{2\cdot (2p)!}
\frac{(4p-2)!}{(2p-1)!}  \beta V_{d-2}~(N'(r_h)-N'(r_c)).
\end{eqnarray}

%%%%%%%%%%%%%%%%%%%%%%%%%%%%%%%%%%%%%%%%%%%%%%
%\subsubsection{Type I instantons in $d=4p$ on Reissner--Nordstr\"om backgrounds}
\subsubsection{Type I instantons in deformed $d=4p$ $p-$Schwarzschild backgrounds}
%%%%%%%%%%%%%%%%%%%%%%%%%%%%%%%%%%%%%%%%%%%%%%%%%

Interestingly, in addition to these solutions given in closed form, we have
constructed numerical solutions with similar properties in other $d=4p$ backgrounds with a
vanishing cosmological constant.
%These are also YM selfdual solutions on other
The only restriction we impose on these backgrounds is to present the expansion
(\ref{m-eh}) as $r \to r_h$ and
to approach asymptotically the $p-$Schwarzschild solution in $p$-Einstein gravity  ($e.g.$
$N(r)\to 1-(r_0/r)^{(2p-1)/p}$ as $r \to \infty$).

These YM instanton solutions have the following expansion near the event horizon\footnote{Note that
for the Schwarzschild-like coordinates we use, the
slope of $w(r)$ diverges as $r \to r_h$. One can easily
verify that this divergence disappears when using instead an isotropic coordinate system.}
\begin{eqnarray}
%\label{ehII}
&&w(r)=w_1\sqrt{r-r_h}+\frac{w_1}{2}
\left(\frac{2(1-2p)-N_2r_h^2}{N_1r_h^2}
-\frac{\sigma_1}{\sigma_0}+w_1^2(p-1)\right)(r-r_h)^{3/2}+o(r-r_h)^{5/2},
%\\
\nonumber\\
&&u(r)=-\frac{N_1\sigma_0}{2}
+\sigma_0(\frac{2p-1}{r_h^2}-\frac{1}{2}N_1(p-1)w_1^2)(r-r_h)+o(r-r_h)^2\label{ehII}
\end{eqnarray}
and at infinity,
\begin{eqnarray}
\label{asII}
w(r)=1-\frac{1}{2}\left(\frac{r_0}{r}\right)^{(2p-1)/p}+\dots,~~
u(r)=-\frac{2p-1}{2pr}\left(\frac{r_0}{r}\right)^{(2p-1)/p}+\dots,
\end{eqnarray}
Our numerical constructions of Type I $p-$YM selfdual solutions is limited here
to those on Reissner-Nordstr\"om $p-$Einstein gravity backgrounds, as interesting examples of the generic case.
This $p-$Reissner--Nordstr\"om metric is parametrised explicitly by the functions~\cite{Chakrabarti:2001di}
\be
\label{t78}
N(r)=1-\bigg[\bigg(\frac{r_0}{r}\bigg)^{2p-1}+\frac{c_2}{r^{2(3p-2)}}\bigg]^{1/p},
\qquad \si(r)=1\,,
\ee
where $r_0>0$ and $c_2\neq0$ is an unspecified constant related to the electric charge,
so that $N(r)$ has exactly one positive root at some $r=r_h$
and $N(r)>0$ for all $r>r_h$. The metric function
\re{t78} pertains to the $p-$Einstein(--Maxwell) system,
which we here refer to as $p-$Reissner--Nordstr\"om.
For small $c_2$, these can be viewed locally as deformations of the
$p-$Schwarzschild double-selfdual backgrounds, which
may give an heuristic explanation for the existence of these YM instantons.

Here we have excluded backgrounds of gravity with
cosmological constant for purely practical reasons. Moreover, while in the presence
of a cosmological constant the Einstein equations are satisfied by a double-selfdual metric,
in the presence of a $U(1)$ field this is not the case. Thus, we learn
something new by employing a $p-$Einstein Reissner-Nordstr\"om background, namely that even when the
background Riemann curvature is not double-selfdual, the YM instantons remain selfdual. While this is consistent
with the assertions in \cite{Charap:1977ww} and \cite{O'Brien:1988rs},
namely that to construct (analytically or numerically) a single-selfdual YM solution it is {\it sufficient}
to employ the embedding \re{emb} of a double-selfdual Riemann curvature, it is not actually {\it necessary}.

For all considered solutions, the gauge functions $w(r)$ and $u(r)$ interpolate monotonically
between the corresponding values at $r=r_h$ and the asymptotic values at infinity,
without presenting any local extrema. Type I solutions exist for all values of the parameter $r_h$, in
contrast to the Type II solutions presented in {\bf 3.2} below, which exist for $r_h$ up to
a maximal value. In Figure 1 we plot the $p=2$ Charap-Duff solution (with $c_2=0$), known in
closed form, together with the numerically numerically evaluated profiles of a typical selfdual YM solution in a
$d=8$ Euclideanised non--double-selfdual background. This last has been chosen to be the $p=2$
Reissner-Nordstr\"om background (with $c_2=0.01$).

Another interesting property of numerically constructed solutions in $d=4p$ with Type I
boundary conditions concerns the solutions to the second order equations rather than the first order
selfduality equations. In this case one might have expected that higher node radial excitations of
the spherically symmetric selfdual solutions existed.
Our numerical results indicate, quite definitely, that no such solutions exist. Had such non--selfdual
solutions, describing the backreaction from gravity on the YM field been found, they
would have been expected to be sphaleron-like configurations.
%There appear
%to exist no such solutions.

For $c_2\neq 0$, the instanton solutions are evaluated numerically.
For any choice of the metric functions
($N(r),~\sigma(r)$) the action of the selfdual solutions
satisfying (\ref{ehII}), (\ref{asII})
is still given by (\ref{action-II}).
An existence proof for Type I solutions
in a general metric background satisfying a suitable set of conditions
is given in the next section.
One can easily verify that the metric functions (\ref{t78}) satisfy the conditions
there.

%%%%%%%%%%%%%%%%%%%%%%%%
\begin{figure}[h!]
\parbox{\textwidth}
{\centerline{
\mbox{
\epsfysize=10.0cm
\includegraphics[width=92mm,angle=0,keepaspectratio]{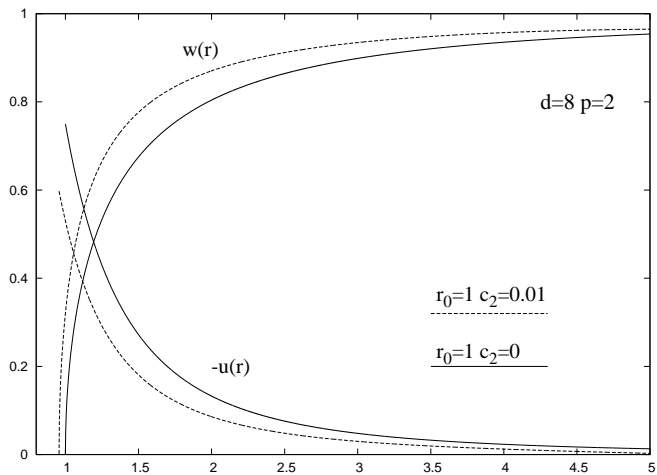}}}}
\caption{{\small The YM gauge functions
are shown as a function of the radial coordinate
$r$  for two Type I $p=2$ YM selfdual solutions, showing the deformation of the explicit DC
solution with $c_2=0$ by the $c_2=0.01$ $2-$Reissner-Nordstr\"om metric}}
\end{figure}
Somewhat surprisingly, it turns out that similar selfdual solutions to $p-$YM systems on {\it other}
spherically symmetric background also exist.
Therefore the condition for the metric background to approach asymptotically
the $p-$Schwarzschild solution in $p$-Einstein gravity is not really crucial after all.
These instanton configurations satisfy the same set of boundary conditions as the
solutions above ($e.g.$ $w(r_h)=0$, $u(r_h)=0$), with an
expansion differs completely from (\ref{ehII}), (\ref{asII}), however.
We have tested this for the example of the $d=8$ $2-$YM system
on the $1-$Reissner--Nordstr\"om background (of the usual $p=1$ Einstein-Maxwell gravity).
These last differ from the former only quantitatively, the
profiles of the functions asymptoting at least one order of magnitude longer, and exhibiting a
similarly magnified steepness at the origin.
Such solutions cannot be viewed as deformations of the CD configurations. However,
the numerical results are supported by the existence proof given in the next section.
A discussion of these more general solutions will be presented elsewhere.

%%%%%%%%%%%%%%%%%%%%%%%%%%%%%%%%%%%%%%%%%%%%%%
\subsubsection{Selfdual Type I Yang-Mills solutions in $d=2(p+q)$}
%%%%%%%%%%%%%%%%%%%%%%%%%%%%%%%%%%%%%%%%%%%%%%%%%
The extension of the Charap-Duff solution in $d=4$ to $d=2(p+q)$ dimensions is also given formally in
\cite{O'Brien:1988rs}. In this case however there exist no black hole solutions, and the only concrete
instantons are those on the symmetric dS/AdS spaces given below.

On flat space the selfduality equations in $d=2(p+q)$ dimensions \re{pq1a}-\re{pq2b}
have no nontrivial solutions, but on a curved spacetime it is possible to
find nontrivial solutions,
albeit on maximally symmetric spaces. These solution minimise absolutely the action of the
reduced one dimensional YM Lagrangian \re{lagpq}.

The dimensions of $\tau_p$ being different from the dimensions of $\tau_q$,
the system \re{lagpq}
is not scale invariant, and the selfduality equations \re{pq1a}-\re{pq2b}
feature the dimensional constant
$\frac{\tau_p}{\tau_q}$, as a result of which no asymptotically flat
solutions to the latter exist.

Unfortunately, the only solution of these equations we could find is
\begin{eqnarray}
w(r)=-\epsilon\sqrt{N(r)},~~u(r)=\epsilon\frac{1}{2}N'(r)\,,
\end{eqnarray}
where
\begin{eqnarray}
\label{bck1}
N=1+\epsilon (\hat\tau_p/\hat\tau_q)^{1/(q-p)}r^2\quad,\quad\sigma(r)=1\,~,
\end{eqnarray}
and $q-p=2n+1$, with $n$ an integer, while $\epsilon=\pm 1$. For  $q-p=2n$ one finds
\begin{eqnarray}
w(r)= \sqrt{N(r)},~~u(r)=-\frac{1}{2}N'(r)\,,
\end{eqnarray}
with the metric functions given by (\ref{bck1}) above.
Restricting for simplicity to the case $\tau_p>0,~\tau_q>0$, we see from \re{bck1} that these
selfdual $(p,q)-$YM instantons are given on an
Euclideanised dS ($\epsilon=-1$) or AdS ($\epsilon=1$) background,
the cosmological constant here being fixed by the coupling constants of the YM model\footnote{
In {\bf 3.1.1} above, where $p=q$,
%pertaining to Type I solutions in $4p$ dimensions,
we studied numerically the second order equations to find out whether there existed
any radial excitations, and the outcome was negative.
Here too we inquire whether there might be non--selfdual solutions
with the matter field deforming the geometry, and found that no such solutions
can exist. We concluded this analytically, by noticing the impossibility to
write for $d=2(p+q)$ (with $p\neq q$) a consistent expansion near $r=r_h$ of the form (\ref{ehII}).}.

%%%%%%%%%%%%%%%%%%%%%%%%%%%%%%%%%%%%%%%%%%%%%%%%%%%%%%%%%%%%%%%%%%%%%%%%%%%
\subsection{Type II solutions: Deformed $p-$Prasad-Sommerfield \\ configurations}
%%%%%%%%%%%%%%%%%%%%%%%%%%%%%%%%%%%%%%%%%%%%%%%%%%%%%%%%%%%%%%%%%%%%%%%%%%%
This subsection deals only with $d=4p$ solutions, and {\bf not} $d=2(p+q)$ ones with $p\neq q$.
The reason for this is that the radial function $w(r)$ in \re{YMsph} in this case vanishes
asymptotically. The scaling properties consistent with finite action require that in $d=2(p+q)$
both $F(2p)$ and $F(2q)$ terms be present in the Lagrangian. Then $w(r)\to 0$ for $r\to\infty$ causes the
contribution of the $F(2p)$ (for $p<q$) to the action to diverge.

The hierarchy of Type II instantons basically consists of the
deformed hierarchy of Prasad--Sommerfield~\cite{Prasad:1975kr} (PS)
monopoles in $4p-1$ dimensions presented in \cite{Radu:2005rf} (cf.
\cite{4,5,6} for analytic proofs of existence and uniqueness of
solutions), generalising the usual $3+1$ dimensional PS
monopoles~\cite{Prasad:1975kr} to $(4p-1)+1$ dimensions.

These $p-$PS monopoles are deformed by the usual ($p=1$) Einstein--Hilbert gravity. We shall
refer to these as $p-$PS monopoles. Here we have used
only $p=1$ gravity in all $4p$ dimensions, since the background gravity here does not play a special
role as it does in the Type I cases. It would have been equally valid to employ any $p-$Einstein gravity
instead, but we chose to work with the simplest background. Type II instantons
%have different properties as compared
differ substantially from the Type I solutions given in {\bf 3.1}. In particular, they satisfy a
different set of boundary conditions and have different actions.

These solutions are found for a set of boundary conditions familiar from
previous studies on gravitating
non Abelian solutions possesing an event horizon, where the YM connection
$A_{\mu}$ has a nonvanishing electric
component $A_0$ (see e.g. \cite{Brihaye:1998cm}, \cite{Bjoraker:2000qd}).
Here the magnetic gauge potential $w$ starts from a nonzero value at the
horizon and vanishes at infinity,
while the electric one $u$ behaves in the opposite way.
%%%%%%%%%%%%%%%%%%%%%%%%
\begin{figure}[h!]
\parbox{\textwidth}
{\centerline{
\mbox{
\epsfysize=10.0cm
\includegraphics[width=92mm,angle=0,keepaspectratio]{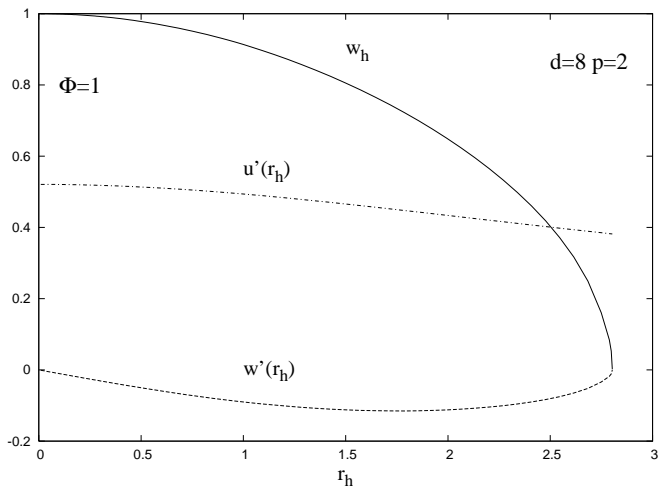}
}}}
\caption{{\small The parameters $w_h,~w'(r_h)$ and $u'(r_h)$
of the $p=2$ type II YM instantons in a $d=8$ Schwarzschild-Tangherlini background
are plotted as a function of $r_h$.}}
\end{figure}
%%%%%%%%%%%%%%%%%%%%%%%%
%
%
The YM potentials have the following expansion as $r \to r_h$
\begin{eqnarray}
w(r)&=&w_h+\frac{(2p-1)w_h(w_h^2-1)}{r_h^2 N_1}(r-r_h)+o(r-r_h)^2\,,\nonumber\\
u(r)&=&\frac{(2p-1)\sigma_h(1-w_h^2)}{r_h^2}(r-r_h)+o(r-r_h)^2\label{ehI}\,,
\end{eqnarray}
%%%%%%%%%%%%%%%%%%%%%%%%
with $0\leq w_h\leq 1$.
\\
For large $r$, the solution reads
\begin{eqnarray}
&&w(r)=\frac{e^{-\Phi r}}{r^{\Phi r_0-1}}+\dots,~~{\rm for}~p=1,~~
w(r)=r^{2p-1}  e^{-\Phi(r- (\frac{r_0}{r})^k/(k-1)) }  +\dots,~~{\rm for}~p\neq 1,~~
\nonumber\\
&&{\rm ~~~~~~~and}~~~u(r)= \Phi-\frac{(2p-1)}{r}+\dots,~~{\rm for~any}~p~,\label{infI}
\end{eqnarray}
where $\Phi$ is an arbitrary nonzero constant.
%\footnote{By using
%a suitable rescaling, one can set $\Phi=1$ without any loss of generality.}
From (\ref{action-p}), we find the action of the Type II instanton
solutions
\begin{eqnarray}
\label{action-I}
S= \frac{\tau_p}{2\cdot (2p)!}\frac{(4p-2)!}{(2p-1)!}\,\beta\, V_{d-2}\,\Phi~.
\end{eqnarray}
One can see that the properties of the background metric enter here through
the expression of $\beta$   ---   the periodicity of the
Euclidean time coordinate. Employing \re{ehI} and \re{infI} to estimate the intergral
of (\ref{4psd2}), implies the existence of a maximal allowed magnitude
of the electric potential at infinity for a given $r_h$
\begin{eqnarray}
\label{cond}
\Phi< (2p-1)\int_{r_h}^\infty dr~\frac{\sigma(r)}{r^2}.
\end{eqnarray}
In practice, we choose $\Phi=1$ without
any loss of generality, which sets the maximal value of the $r_h$ for a given background.
This is in contrast to the Type I solutions where the value of the
horizon radius $r_h$ is not constrained.

%%%%%%%%%%%%%%%%%%%%%%%%
\begin{figure}[h!]
\parbox{\textwidth}
{\centerline{
\mbox{
\epsfysize=10.0cm
\includegraphics[width=92mm,angle=0,keepaspectratio]{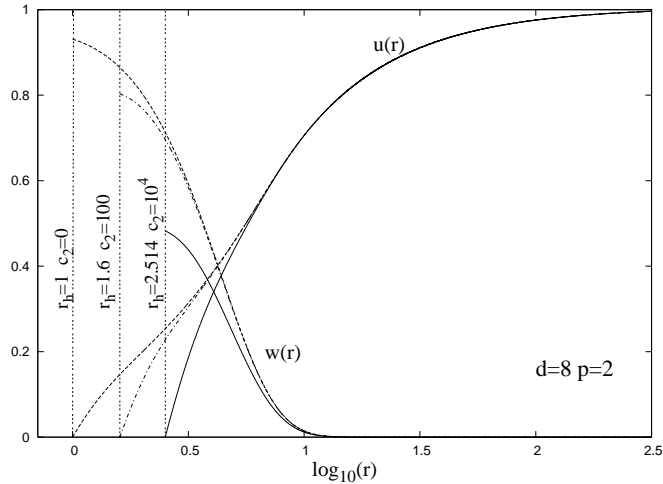}
}}}
\caption{{\small The YM gauge functions are shown as a function of the radial coordinate
$r$ for $p=2$ type II YM selfdual solutions in $d=8$ Euclideanised
Reissner-Nordstr\"om-Tangherlini backgrounds.}}
\end{figure}
%%%%%%%%%%%%%%%%%%%%%%%%
%
In Ref. \cite{Brihaye:2006bk}, arguments for the existence of $p=1$ type II
selfdual Yang-Mills instantons for several $d=4$ spherically symmetric
backgrounds with Euclidean signature were presented. These solutions
were evaluated numerically.
The existence of similar solutions for any nonextremal $SO(3)$-spherically symmetric background
approaching at infinity the $d=4$ Euclideanised Minkowski spacetime was also conjectured.
These solutions can be interpreted as curved spacetime
deformations of the well known Prasad-Sommerfield (PS) monopoles~\cite{Prasad:1975kr},
viewed as instantons of the YM theory in a $R^3\times S^1$ background. Here we extend some of
these arguments to the $p-$PS monopoles~\cite{Radu:2005rf} in $4p-1$ spacelike dimensions.
It is natural to suppose that these solutions will survive
when the background has a nontrivial geometry, at least for a small curvature. Unlike in the previous case
of Type I instantons where the gravitational background was specified to be the $p-$Einstein Schwarzschild like
solution (with or without cosmological constant), here the curving of the background is not similarly
constrained. Like in the $4$ dimensional case \cite{Brihaye:2006bk}, here
we have found numerical arguments for the existence of an hierarchy of $d=4p$ YM selfdual solutions,
the $p=1$ case in \cite{Brihaye:2006bk} being the first member only.
For $p=2,~3$, we have considered several different spherically symmetric backgrounds, the
Schwarzschild-Tangherlini and  Reissner-Nordstr\"om-Tangherlini solutions
in Einstein-Maxwell theory being the simplest cases
(the metric functions for the second situation are
$\sigma=1,~N=1-(\left(\frac{r_0}{r}\right)^{d-3}+\frac{c_2^2}{r^{2(d-3)}})$,
$r_0$ being related to the mass and
$c_2$ to the electric charge of the fixed backgrounds, respectively).

As in the $p=1$ case, the $r_h \to 0$ limit (when physically possible), provides
instanton solutions in a topologically trivial background.
This is nicely illustrated by the case of $p-$YM selfdual instantons
in the background of a Einstein-Yang-Mills purely magnetic
hairy black holes discussed in \cite{Brihaye:2002hr},
which solve also the field equations for an Euclidean signature.
These solutions have a particle-like globally-regular limit
with a nonvanishing curvature, the Killing vector
$\partial/\partial \tau$ presenting in this case no fixed point sets
($i.e.$ $g_{\tau \tau}>0$ for any $r\geq 0$ and an arbitrary periodicity $\beta$).
When taking instead the $r_h\to 0$ limit for
a Schwarzschild background, the $p-$PS-type configurations in \cite{Radu:2005rf} are approached.
In all these cases, the approximate expression of the YM instanton solutions as $r \to 0$ is
\begin{eqnarray}
\label{origin}
w(r)=1-br^2+O(r^4),~~
u(r)= 2b\sigma_0 r+O(r^2),
\end{eqnarray}
(with $b>0$ and $\sigma_0=\sigma(r=0)$),
the asymptotic form (\ref{infI}) being valid in this case, too.

In all cases, the gauge functions $w(r)$ and $u(r)$ interpolate monotonically
between the corresponding values at $r=r_h$ and the asymptotic values at infinity,
without presenting any local extrema. For small enough values of $r_h$, the solutions
look very similar to the flat space selfdual YM configuration.
These solutions get deformed with the value of $r_h$ increasing, while the value
of the magnetic potential $w$ at $r=r_h$ steadly decreases.
As $r_h$ approaches some maximal value implied by (\ref{cond}), we find that $w_h\to 0$ and the
solution approaches the limiting configuration
\begin{eqnarray}
\label{ab}
w(r)=0,~~u(r)=\Phi+(2p-1)\int \frac{\sigma(r)}{r^2}~dr.
\end{eqnarray}

In Figures 2 we plotted several relevant parameters of the
YM instanton solutions as a function of $r_h$, for the case of $p=2~(d=8)$
Schwarzschild-Tangherlini background. A typical selfdual YM solution in a
Euclideanised  Reissner-Nordstr\"om-Tangherlini black hole is plotted in Figure 3.
These plots retain the generic features of the picture we found in other cases.

%%%%%%%%%%%%%%%%%%%%%%%%%%%%%%%%%%%%%%%%%%%%%%
\section{Analytic proofs of existence}
\setcounter{equation}{0}
%\setcounter{0}
%%%%%%%%%%%%%%%%%%%%%%%%%%%%%%%%%%%%%%%%%%%%%

To underpin the numerically constructed solutions with spherical symmetry in $d-1$ dimensions
presented in the previous section, we present analytic existence proofs for these in the present section.
As in section {\bf 3} above, we have split this section into two subsections, dealing with Types I
and II solutions respectively.

%%%%%%%%%%%%%%%%%%%%%%%%%%%%%%%%%%%%%%%%%%%%%
\subsection{Type I Solutions}
\subsubsection{The problem}
%%%%%%%%%%%%%%%%%%%%%%%%%%%%%%%%%%%%%%%%%%%%%
In this subsection, we present an analytic proof for the existence of
those Type I solutions which
were evaluated numerically in the previous section.
%The explict metric used in this case was the
%$p-$Reissner-Nordstr\"om background in the absence of a cosmological constant, which is parametrised
%explicitly in terms of the metric functions given by \re{t78}.

%In our applications, the metric function $N$ may be chosen to take the explicit form
%\be\label{t78}
%N(r)=1-\bigg[\bigg(\frac{r_0}{r}\bigg)^{2p-1}+\frac c{r^{2(3p-2)}}\bigg]^{1/p},
%\ee
%where $r_0>0$ and $c\neq0$ is an unspecified constant so that $N(r)$ has exactly one positive root at some
%$r=r_h$ and $N(r)>0$ for all $r>r_h$.

Without loss of generality, we will consider only the case of upper signs in
the system of selfduality equations \re{4psd1} and \re{4psd2} over $(r_h,\infty)$,
%\bea
%\sqrt{\sigma N} w'+\frac1{\sqrt{\sigma N}} uw&=&0, \quad r_h<r<\infty,\label{t51}\\
%\frac r{\sqrt{\sigma(2p-1)}}[(1-w^2)^{p-1} u]'-\frac{\sqrt{\sigma(2p-1)}}r(1-w^2)^p&=&0,\quad
%r_h<r<\infty,\label{t52}
%\eea
subject to the boundary conditions
\bea
w(r_h)&=&0,\quad u(r_h)=u_h,\label{t1.4}\\
w(\infty)&=&1,\quad u(\infty)=0,\label{t1.5}
\eea
where $u_h<0$ is a constant. We will be interested in solutions such that $u$ remains nonpositive and $w$ remains
nonnegative for all $r>r_h$.

From (\ref{4psd1}), we have \be \label{t1.6} \sigma N w'+wu=0,\quad
r>r_h, \ee which implies that $w'\geq0$ for all $r>r_h$. In fact, we
also have $w>0$ everywhere. Indeed, if there is an $r_0>r_h$ such
that $w(r_0)=0$, then $w\equiv0$ due to the uniqueness theorem for
initial value problems of ordinary differential equations, which
violates the boundary condition for $w$ stated in (\ref{t1.5}).
Similarly, $w<1$ everywhere. Otherwise, if there is an $r_0>r_h$
such that $w(r_0)=1$, then $w(r)\equiv1$ for all $r\geq r_0$. Using
the analyticity of solutions in the BPS system of equations,
(\ref{4psd1}) and (\ref{4psd2}), we see that $w(r)\equiv1$ for all
$r>r_h$, which contradicts $w(r_h)=0$ in (\ref{t1.4}). These
established facts now allow us to assert that $u(r)<0$ for all
$r>r_h$. Suppose otherwise that there is an $r_0>r_h$ such that
$u(r_0)=0$. Hence, $r_0$ is a maximum point for $u$ and $u'(r_0)=0$.
Inserting these into (\ref{4psd2}) evaluated at $r=r_0$, we obtain
$w(r_0)=1$, which is false.  A special consequence from the
conclusion $w>0, u<0$ and (\ref{t1.6}) is that $w'>0$ for all
$r>r_h$. Another is that the fact $w>0$ allows us to suppress
(\ref{t1.6}) into \be \label{t1.7}
%\label{t53}
-\sigma N (\ln w)'=u,\quad r>r_h.
\ee

Inserting (\ref{t1.7}) into (\ref{4psd2}), we have
\be\label{t1.8}
[(1-w^2)^{p-1}\sigma N (\ln w)']'+\frac{(2p-1)\sigma}{r^2} (1-w^2)^p=0,\quad r>r_h.
\ee
Furthermore, with $v=\ln w$ or $w=\e^v$, we rewrite (\ref{t1.8}) into the form
\be\label{t1.9}
[(1-\e^{2v})^{p-1}\sigma N v']'+\frac{(2p-1)\sigma}{r^2} (1-\e^{2v})^p=0,\quad r>r_h,
\ee
so that the boundary condition for $w$ is converted to the boundary condition for $v$ which says
\be\label{t1.10}
v(r_h)=-\infty,\quad v(\infty)=0.
\ee
Recall that since $w(r)$ stays within the interval $(0,1)$ when $r> r_h$, the range of $v(r)$ for
$r>r_h$ is $(-\infty,0)$. This property suggests that
we may use the invertible transformation from $(-\infty,0]$ to itself defined by
\be
f=P(v)=\int_0^v(1-\e^{2s})^{p-1}\,\dd s.\label{t9}
\ee
to simplify (\ref{t1.9}) further into
\be\label{t54}
f''+f'(\ln[\sigma N])'+\frac{(2p-1)}{r^2 N} (1-\e^{2Q(f)})^p=0,\quad r_h<r<\infty.
\ee
Here and in the sequel, we use $Q$ to denote the inverse of $P$ over $(-\infty,0]$.
It is clear that $P, Q$ are increasing and $P(0)=Q(0)=0$.

%%%%%%%%%%%%%%%%%%%%%%%%%%%%%%%%%%%%%%%%%%%%%
%\subsubsection{A $p=1$ Reissner-Nordst\"om background}
%%%%%%%%%%%%%%%%%%%%%%%%%%%%%%%%%%%%%%%%%%%%%
To motivate our general study,
we start from the simplest (but instructive) situation in (\ref{4psd1}) and
(\ref{4psd2}) for which $p=1$ in
(\ref{t78})
and $\sigma\equiv1$.  Therefore $N(r)$ takes the form
\be\label{t79}
N(r)=1-(\frac{r_0}r+\frac{c_2}{r^2}).
\ee
It is seen that the function $N(r)$ has a single positive root if and only if
$c_2>0$ and the root is given by
\be\label{t80}
r_h=\frac12(r_0+\sqrt{r_0^2+4c_2}),
\ee
which allows us to rewrite (\ref{t79}) as
\be\label{t81}
N(r)=\bigg(1-\frac{r_h}r\bigg)\bigg(1+\frac{R_h}r\bigg)=\frac1{r^2}(r-r_h)(r+R_h)
\ee
for some number $R_h>0$ and consider the equations over $r_h<r<\infty$.

With $\sigma\equiv1$ and $N$ given in (\ref{t81}), the equation (\ref{t54}) becomes
\be\label{t82}
f''+f'\bigg(\frac1\rho+\frac1{\rho+r_h+R_h}-\frac2{\rho+r_h}\bigg)+\frac{1}
{\rho(\rho+r_h+R_h)}(1-\e^{2Q(f)})=0,
\ee
where we have used the translated radial variable $\rho=r-r_h>0$. Using the Euler transformation
 $\rho=\e^t$, we obtain from (\ref{t82}) the
equation
\be\label{t83}
f_{tt}-f_t g(t)+h(t)(1-\e^{2Q(f)})=0,\quad -\infty <t<\infty,
\ee
where the coefficient functions $g(t)$ and $h(t)$ are given by the expressions
\be\label{t84a}
g(t)=\frac{\e^t(\e^t+r_h+2R_h)}{(\e^t+r_h)(\e^t+r_h+R_h)},
\ee
\be \label{t84b}
h(t)=\frac{\e^t}{(\e^t+r_h+R_h)}.
\ee

%%%%%%%%%%%%%%%%%%%%%%%%%%%%%%%%%%%%%%%%%%%%%
\subsubsection{The proof}
%%%%%%%%%%%%%%%%%%%%%%%%%%%%%%%%%%%%%%%%%%%%%
We now consider the general situation when $p\geq1$ and $\sigma$ is
arbitrary. With the same sequence of variable substitutions, we
rewrite the governing equation in terms of the radial variable $r$
as (\ref{t54}).

 Similar to (\ref{t81}), we express $N(r)$ as
\be\label{t87}
N(r)=\bigg(1-\frac{r_h}r\bigg) M(r),
\ee
where $M>0$ for all $r\geq r_h$ and $M(\infty)=1$. With $\rho=r-r_h$, $t=\ln \rho$, and
\bea\label{t88a}
g(t)&=&1-\bigg[\rho\bigg(\frac{\sigma'(r)}{\sigma(r)}+\frac{N'(r)}{N(r)}\bigg)\bigg]_{\rho=\e^t},\\
\quad h(t)&=&\bigg[\frac{(2p-1)\rho}{rM(r)}\bigg]_{\rho=\e^t},\label{t88b}
\eea
the equation (\ref{t54}) becomes
\be\label{t89}
f_{tt}-f_t g(t)+h(t) (1-\e^{2Q(f)})^p=0,
\ee
subject to the updated boundary condition
\be\label{t1.24}
f(-\infty)=-\infty,\quad f(\infty)=0.
\ee
It is seen that (\ref{t89}) generalizes (\ref{t83}). In view of (\ref{t84a}) and (\ref{t84b}), we impose the following conditions
on the coefficient functions $g(t)$ and $h(t)$:

(i) $g(t)\geq0, h(t)>0$ for all $t$;

(ii) as $t\to -\infty$, there are the asymptotics
\be\label{t90}
g(t)=\mbox{O}(\e^{\delta t}),\quad h(t)=\mbox{O}(\e^{\vep t})
\ee
for some constant $\delta, \vep>0$;

(iii) there hold
\be\label{t91}
\lim_{t\to\infty} g(t)=g(\infty)>0,\quad \lim_{t\to\infty} h(t)=h(\infty)>0.
\ee
Note that
in view of the definition (\ref{t88a})--(\ref{t88b}) and the fact that
$M(\infty)=1$ we actually have $h(\infty)=2p-1$.
However, this precise value is not important for our subsequent discussion.

In order to solve (\ref{t89}) subject to (\ref{t1.24}), we
consider the solution of the equation (\ref{t89}) over the interval $-\infty<t<\infty$ subject to the initial value condition
\be\label{t93}
f(t_0)=-a,\quad f_t(t_0)=b,
\ee
where $t_0\in(-\infty,\infty)$ and $a>0$. We shall show that for any $a>0$, there exists a unique
number $b(a)>0$ so that when $b=b(a)$ the initial value problem
consisted of (\ref{t89}) and (\ref{t93}) has a uniquely and globally
defined solution $f(t)$ satisfies $f_t(t)>0$ and $f(t)<0$ for all $t$. Moreover, such a solution
satisfies the desired boundary condition (\ref{t1.24}).

For technical reasons, we shall also consider the possibility that the solution $f$
of (\ref{t89}) and (\ref{t93}) takes positive values under certain
initial conditions. Consequently we need to modify (\ref{t89}) as
\be\label{t1.28}
f_{tt}-f_t g(t)=h(t) R(f),
\ee
where we define
\be \label{tR}
R(s)=\left\{\begin{array}{cc} -(1-\e^{2Q(s)})^p,& s\leq0,\\ 2p s,& s>0,\end{array}\right.
\ee
 so that $R(s)$ is a differentiable increasing function.

For given $t_0$ and fixed $a>0$, we use $f(t;b)$ to represent the unique solution of (\ref{t1.28}) satisfying
(\ref{t93}) which is defined in its local or global interval of existence.

We will conduct a shooting analysis. To this end, we define our sets of shooting slopes as follows:
\bea
{\cal S}^-&=&\{b\in\bfR\,|\,\exists t>t_0 \mbox{ so that } f_t (t;b)<0\},\nn\\
{\cal S}^0&=&\{b\in\bfR\,|\,f_t(t;b)>0\mbox{ and } f(t;b)\leq0\mbox{ for all }t>t_0\},\nn\\
{\cal S}^+&=&\{b\in\bfR\,|\,f_t(t;b)>0\mbox{ for all }t\geq t_0\mbox{ and }f(t;b)>0
\mbox{ for some }t>t_0\}.
\nn
\eea

\begin{lemma}\label{lemma1.1}
The set of real numbers $\bfR$ may be expressed as the disjoint union $\bfR={\cal S}^-\cup{\cal S}^0\cup{\cal S}^+$.
\end{lemma}
\begin{proof}
Let $t>t_0$ be any point in the interval of existence of the
solution $f(\cdot;b)$. Of course $(-\infty,0)\subset {\cal S}^-$.
For any $b\not\in {\cal S}^-$, we have $f_t(t;b)\geq0$ for all
$t\geq t_0$. We claim that $f_t(t;b)>0$ everywhere. In fact, if
there is some point $t_1>t_0$ such that $f_t(t_1;b)=0$, then
$f(t_1;b)\neq0$ since $f=0$ is an equilibrium of the equation
(\ref{t1.28}) which cannot be attained by a solution trajectory
originating from a non-equilibrium initial state. Using the fact
that $f_t(t_1;b)=0$ but $f(t_1;b)\neq 0$ in (\ref{t1.28}), we have
\be\label{t1.30} f_{tt}(t_1;b)=h(t_1)R(f(t_1;b))\neq0. \ee Hence,
depending on the sign of $f_{tt}(t_1;b)$, we have either
$f_t(t;b)<0$ for $t<t_1$ but $t$ is close to $t_1$ when
$f_{tt}(t_1;b)>0$ or $f_t(t_1;b)<0$ for $t>t_1$ but $t$ is close to
$t_1$ when $f_{tt}(t_1;b)<0$. Therefore, $b\in {\cal S}^-$, a
contradiction. Hence $f_t(t;b)>0$ for all $t>t_0$ which proves
$b\in{\cal S}^-\cup {\cal S}^+$ as claimed.
\end{proof}

\begin{lemma}\label{lemma1.2}
The sets ${\cal S}^-$ and ${\cal S}^+$ are both open and nonempty.
\end{lemma}

\begin{proof} The set ${\cal S}^-$ is of course nonempty because $(-\infty,0)\subset
{\cal S}^-$ by the definition of ${\cal S}^-$. The openness of ${\cal S}^-$ follows immediately from the continuous dependence theorem of
the solution of an ordinary differential equation on its initial values.

We now prove that ${\cal S}^+$ is also nonempty. To this end, we observe that, when $b>0$, $f_t(t;b)$ remains
positive for $t\in(t_0, t_0+\vep)$ when $\vep>0$ is small enough. Since $g(t)\geq0$, we see that (\ref{t1.28}) gives
us $f_{tt}\geq h(t)R(f)$. Integrating this inequality twice and using the initial condition (\ref{t93}), we have
\bea
f_t(t;b)&\geq& b +\int_{t_0}^t h(s_1) R(f(s_1;b))\,\dd s_1,\quad t_0<t<t_0+\vep,\label{t1.31}\\
f(t;b)&\geq &-a +b(t-t_0) \nn\\
&& +\int_{t_0}^t\int_{t_0}^{s_2} h(s_1)R(f(s_1;b))\,\dd s_1\,\dd s_2,\quad t_0<t<t_0+\vep.\label{t1.32}
\eea
Of course, (\ref{t1.31}) and (\ref{t1.32}) continue to hold wherever $f_t(t;b)\geq0$ ($t>t_0$). We show that, when $b>0$ is large enough, we have $b\in{\cal B}^+$. In fact, for any $t_1>t_0$, the slope number $b>0$ can be chosen
so that
\bea
&& b +\int_{t_0}^{t_1} h(s_1) R(-a)\,\dd s_1>0,\label{t1.33}\\
&&-a +b(t_1-t_0)
+\int_{t_0}^{t_1}\int_{t_0}^{s_2} h(s_1)R(-a)\,\dd s_1\,\dd s_2>0.\label{t1.34}
\eea
Initially, since $f_t(t;b)>0$, we have $f(t;b)>f(t_0;b)=-a$ (for $t>t_0$). Hence $R(f(t;b))>R(-a)$. In view of
(\ref{t1.31}) and (\ref{t1.33}), we get
\be
f_t(t;b)>b +\int_{t_0}^{t_1} h(s_1) R(-a)\,\dd s_1>0,\quad t_0<t\leq t_1,
\ee
which implies $f(t;b)>f(t_0;b)=-a$ and $R(f(t;b))>R(-a)$ for all $t_0<t\leq t_1$.  Using this fact in
(\ref{t1.32}) and (\ref{t1.34}), we have $f(t_1;b)>0$. Since $f(t;b)$ strictly increases in $(t_0,t_1)$,
there is a unique point $t_2\in (t_0,t_1)$ such that $f(t_2;b)=0$ but $f(t;b)<0$ for all $t\in(t_0,t_2)$.
However, the definition (\ref{tR}) says that $R(f)\geq0$ whenever $f\geq0$, we see that for all $t>t_2$, we have
\bea
f_t(t;b)&\geq& b +\int_{t_0}^t h(s_1) R(f(s_1;b))\,\dd s_1\geq  b +\int_{t_0}^{t_2} h(s_1) R(f(s_1;b))\,\dd s_1\nn\\
&\geq& b +\int_{t_0}^{t_1} h(s_1) R(-a)\,\dd s_1>0,\label{t1.36}\\
f(t;b)&>&0,
\eea
which establishes $b\in{\cal S}^+$ and the nonemptyness of ${\cal S}^+$ follows.

It is not hard to show that ${\cal S}^+$ is open. In fact, let
$b_0\in {\cal S}^+$. Then $f_t(t;b_0)>0$ for all $t>t_0$ and there
is a $t_3>t_0$ so that $f(t_3;b_0)>0$. By the continuous dependence
theorem for the solution to the initial value problem of an ordinary
differential equation, we see that when $b$ is sufficiently close to
$b_0$, we still have $f(t_3;b)>0$ and $f_t(t;b)>0$ for all $t\in
[t_0, t_3]$. Applying the same argument as that for deriving
(\ref{t1.36}), we conclude that $f_t(t;b)>0$ for all $t>t_3$ as
well. Therefore $b\in {\cal S}^+$ and ${\cal S}^+$ is indeed open.
\end{proof}

\begin{lemma}\label{lemma1.3}
The set ${\cal S}^0$ is nonempty and closed. Furthermore, for $b\in {\cal S}^0$, we have $f(t;b)<0$ for all $t>t_0$ and
$f(t;b)\to0$ as $t\to\infty$.
\end{lemma}
\begin{proof}
Since $\bfR$ is connected, it cannot be expressed as the disjoint union of two open sets, ${\cal S}^-$ and ${\cal S}^+$
established in Lemma \ref{lemma1.2}. Hence ${\cal S}^0$ is nonempty and closed.

The definition of ${\cal S}^0$
gives us $f(t;b)\leq0$ for all $t>t_0$. If there is a point $t_1>t_0$ such that $f(t_1;b)=0$, then $f_t(t_1;b)=0$
which is false.

Since $f(t;b)$ increases and stays negative-valued for all $t>t_0$, the limit
\be \label{t1.38}
\eta\equiv\lim_{t\to\infty} f(t;b)
\ee
exists and satisfies $-a<\eta\leq0$. The finiteness of the limit $\eta$ in (\ref{t1.38}) implies that there is
a sequence $\{t_j\}$ ($t_j\to\infty$ as $j\to\infty$) so that
\be\label{t1.39}
f_t(t_j;b)\to0\quad\mbox{ as }j\to\infty.
\ee
As a consequence
of (\ref{t1.39}) and (\ref{t91}),
we see from (\ref{t1.28}) (or more precisely, (\ref{t89})) that $f_{tt}(t;b)$ will stay below a negative number when $t$ is sufficiently large if $\eta<0$,
which contradicts (\ref{t1.39}).
\end{proof}

In fact, we know more about the `good slope' set ${\cal S}^0$. We have

\begin{lemma}\label{lemma1.4}
The set ${\cal S}^0$ consists of a single point.
\end{lemma}
\begin{proof}
Let $b_1$ and $b_2$ be taken from ${\cal S}^0$ and $f(t;b_1)$ and $f(t;b_2)$ the corresponding solutions of (\ref{t89})
and (\ref{t93}). Then $z(t)=f(t;b_1)-f(t;b_2)$ ($t\geq t_0$) satisfies
\be\label{t1.40}
z_{tt}-g(t)z_t =h(t)R'(\xi(t))z,\quad t_0<t<\infty,
\ee
where $\xi(t)$ lies between the quantities $f(t;b_1)$ and $f(t;b_2)$. Using $h(t)R'(\xi(t))>0$, $z(t_0)=z(\infty)=0$,
and the maximum principle in (\ref{t1.40}), we deduce $z\equiv0$. In particular, $z_t(t_0)=b_1-b_2=0$ as
claimed.
\end{proof}

In view of Lemma \ref{lemma1.4}, for given $a>0$ in (\ref{t93}), let the unique point in ${\cal S}^0$ be denoted
by $b=b(a)$ and the corresponding solution of (\ref{t89}) and (\ref{t93}) be simply denoted by $f=f(t)$. We have

\begin{lemma}\label{lemma1.5}
For $b=b(a)$, the solution $f(t)$ of (\ref{t89}) and (\ref{t93}) exists globally for all $t$.
Furthermore, it satisfies $f_t(t)>0$
and $f(t)<0$ for all $t$ and realizes the other expected boundary condition
\be\label{t1.41}
\lim_{t\to-\infty} f(t)=-\infty.
\ee
\end{lemma}
\begin{proof}
With the notation just mentioned, we consider the solution over the
left-half line $t<t_0$. Multiplying (\ref{t89}) by $\e^{\int_t^{t_0}
g(s)\,\tiny{\dd} s}$ and integrating, we get \be\label{t95}
f_t(t)=\e^{-\int_t^{t_0} g(s)\tiny{\dd} s}\bigg(b(a)+\int_t^{t_0}
h(s) (1-\e^{2Q(f(s))})^p\e^{\int_s^{t_0} g(s_1)\tiny{\dd}s_1}\,\dd
s\bigg),\quad t<t_0. \ee In particular, \be \label{t98a}
f_t(t)>b(a)\e^{-\int_t^{t_0} g(s)\tiny{\dd} s}\geq
b(a)\e^{-\int_{-\infty}^{t_0} g(s)\tiny{\dd} s}\equiv b_0>0 \ee for
all $t<t_0$, where the convergence of the improper integral in
(\ref{t98a}) follows from (\ref{t90}). So $f(t)<f(t_0)-b_0(t_0-t)$
for $t<t_0$ and we obtain $f(-\infty)=-\infty$ as claimed.
\end{proof}

We can now check the boundary conditions for the original field configuration pair $w$ and $u$ in terms of the radial
variable $r$.

First, using the relations $v=\ln w$ and $v=Q(f)$, we may immediately deduce from Lemmas \ref{lemma1.3}
and \ref{lemma1.5} that $w(r)\to1$ as $r\to\infty$  and $w(r)\to 0$ as
$r\to r_h$, respectively.

Next, since (by (\ref{t1.7}))
\bea\label{t96}
u&=&-\sigma N\frac{\dd v}{\dd r}=-\frac{\sigma(r)M(r)} r\rho \frac{\dd v}{\dd\rho}\nn\\
&=&-\frac{\sigma(\e^t+r_h)M(\e^t+r_h)}{(\e^t+r_h)}\frac{\dd v}{\dd t}\nn\\
&=&-\frac{\sigma(\e^t+r_h)M(\e^t+r_h)}{(\e^t+r_h)}(1-\e^{2Q(f(t))})^{-(p-1)} f_t(t),
\eea
we can use the statement $f(-\infty)=-\infty$ in Lemma \ref{lemma1.5} to arrive at the expression
\be\label{t97}
\lim_{r\to r_h} u(r)=-\frac{\sigma(r_h)M(r_h)}{r_h} \lim_{t\to-\infty} f_t(t)\equiv -\frac{\sigma(r_h)M(r_h)}{r_h}f_t(-\infty).
\ee
Note that, using (\ref{t95}), we have
\be\label{t95a}
f_t(-\infty)=\e^{-\int_{-\infty}^{t_0} g(s)\tiny{\dd} s}\bigg(b(a)+\int_{-\infty}^{t_0} h(s) (1-\e^{2Q(f(s))})^p\e^{\int_s^{t_0} g(s_1)\tiny{\dd}s_1}\,\dd s\bigg)
\ee
and the uniform convergence of the right-hand side of (\ref{t95a}) is a consequence of the assumption (\ref{t90}).
In particular, the left-hand side of (\ref{t95a}) is a well-defined positive number
which gives rise to the negative limiting value of $u$ at $r=r_h$.

In order to see what happens for $u$ when $r\to \infty$, we can linearize (\ref{t89}) around $t=\infty$ to get
\be
\theta_{tt}-g(\infty)\theta_t -2[h(\infty)p]\theta=0
\ee
 which has exactly one negative characteristic root, $-\lm$ (say).
Therefore $f$
vanishes at $t=\infty$ exponentially fast
like $\e^{-\lm t}$. Using (\ref{t42}), we have $v=Q(f)=\mbox{O}(\e^{-\lm t/p})$ when
$t$ is large. Inserting these results into (\ref{t96}) and noting that $f_t(t)=\mbox{O}(\e^{-\lm t})$ for $t$ large,
we have
\be\label{t991}
u=-\frac{\sigma(\e^t+r_h)M(\e^t+r_h)}{(\e^t+r_h)}\mbox{O}(\e^{\lm(p-1)t/p})  f_t(t)=\mbox{O}(\e^{-\lm t/p})\quad \mbox{as }t\to\infty.
\ee
Therefore we have shown that $u(r)\to 0$ as $r\to\infty$ as expected.

\medskip

Returning to the original variables, we can summarize our study of the Type I solutions as follows.

\begin{theorem}
Suppose that the background metric functions $N(r)$ and $\sigma(r)$ satisfy the conditions
that $N(r)$ has exactly one positive root $r=r_h$ (say), $N(r)>0$ when $r>r_h$,
\be\label{t99}
\lim_{r\to\infty} N(r)\equiv N(\infty)>0, \quad
\ee
\be
\lim_{r\to\infty}\bigg[1-(r-r_h)\bigg(\frac{\sigma'(r)}{\sigma(r)}+\frac{N'(r)}{N(r)}\bigg)\bigg]\equiv g(\infty)>0,
\ee
there are constants $\delta, \vep>0$ such that for $r$ near $r_h$, there holds
\be
1-(r-r_h)\bigg(\frac{\sigma'(r)}{\sigma(r)}+\frac{N'(r)}{N(r)}\bigg)=\mbox{O}((r-r_h)^\delta),\quad
\ee
\be
\label{tnew}
\frac{(r-r_h)^2}{r^2N(r)}=\mbox{O}((r-r_h)^\vep),
\ee
and for all $r>r_h$, there is the bound
\be\label{t101}
(r-r_h)\bigg(\frac{\sigma'(r)}{\sigma(r)}+\frac{N'(r)}{N(r)}\bigg)\leq 1.
\ee
Then the BPS system of equations (\ref{4psd1}) and (\ref{4psd2}) has a solution pair $(w,u)$ over
$r>r_h$ satisfying the boundary condition
\be\label{t102}
w(r_h)=0,\quad w(\infty)=1;\quad u(r_h)=u_h,\quad u(\infty)=0,
\ee
where $u_h<0$ is a suitable constant, $w(r)>0$, $w'(r)>0$, and $u<0$ for all $r>r_h$.
\end{theorem}

It can easily be seen that the conditions (\ref{t99})-(\ref{tnew})
are satisfied by any reasonable metric background and are in
agreement with the asymptotics at the beginning of the section 2.
The requirement (\ref{t101}) appears to be difficult to prove for an
arbitrary metric. However, we have verified that this condition is
satisfied in the concrete case we have considered in the numerics.
%However, it may be checked directly that our conditions for the above existence
%theorem are in line with the concrete
%case where  $N$ and $\sigma$ are given by (\ref{4psd1}).

%%%%%%%%%%%%%%%%%%%%%%%%%%%%%%%%%%%%%%%%%%%%%%%%%%%%%%%%%%%%
\subsection{Type II solutions}
%\setcounter{equation}{0}
%%%%%%%%%%%%%%%%%%%%%%%%%%%%%%%%%%%%%%%%%%%%%%%%%%%%%%%%%%%%
\subsubsection{The proof for a Schwarzschild background}
%%%%%%%%%%%%%%%%%%%%%%%%%%%%%%%%%%%%%%%%%%%%%%%%%%%%%%%%%%%%
We now consider type II solutions considered in section {\bf 3.2}.
As before, we will start from a concrete
situation.

We first set $\sigma\equiv1$ in
the system of selfduality equations \re{4psd1}-\re{4psd2},
%\bea
%\sqrt{N} w' +\frac1{\sqrt{N}} u w&=&0,\quad r_h<r<\infty,\label{t1}\\
%\frac r{\sqrt{2p-1}}[(1-w^2)^{p-1} u]'-\frac{\sqrt{2p-1}}r(1-w^2)^p &=&0,\quad r_h<r<\infty.
%\label{t2}
%\eea
%The boundary conditions are taken to be
and we seek solutions with boundary conditions
\bea
&&w(r_h)=w_h,\quad \quad u(r_h)=0;\label{t3}\\
&& w(\infty)=0,\quad u(\infty)=\Phi,\label{t4}
\eea
where $w_h\in [0,1]$ and $\Phi>0$ are constants.
For convenience, we shall now concentrate on nonnegative-valued solutions.

Like before, some elementary but useful properties of the solutions of
the equation (\ref{4psd1}), (\ref{4psd1})
together with the boundary conditions
(\ref{t3}), (\ref{t4}) may be deduced immediately.
First, note that (\ref{4psd1})  implies that $w'\leq0$. If $w_h=0$ in (\ref{t3}), then it follows from (\ref{t4}) that
$w\equiv0$. Inserting this into (\ref{4psd2})
and using (\ref{t3}), we obtain $u(r)=(2p-1)(r^{-1}_h-r^{-1})$. Hence, in (\ref{t4}), we have
\be
\Phi=\frac{(2p-1)}{r_h}.
\ee
In other
words, the positive constant $\Phi$ in (\ref{t4}) in this trivial solution situation cannot be arbitrary.
For the nontrivial solution situation, we have $0<w_h\leq 1$.
The uniqueness theorem for the
initial value problem of ordinary differential equations implies that
a nontrivial solution $w$
of (\ref{4psd1})  cannot assume zero value at finite $r>r_h$. Hence $w(r)>0$
for all $r>r_h$ which allows us to rewrite (\ref{4psd1})  as
\be
-N(\ln w)'=u,\quad r>r_h.\label{t5}
\ee
Similarly, $w(r)< 1$ for all $r>r_h$. Otherwise, suppose that
there is an $r_0>r_h$ such that $w(r_0)=1$.  Hence
$w_h=1$ and $w(r)=1$ for all $r_h<r<r_0$. Since the solution is
necessarily analytic at $r_0$, we see that $w(r)=1$
for $r$ around $r_0$ which establishes $w(r)=1$ for all $r>r_h$,
contradicting $w(\infty)=0$ in (\ref{t4}).
We assert that $u(r)>0$ for all $r>r_h$. Otherwise, suppose there is an $r_0>r_h$
such that $u(r_0)=0$. Then
$u$ attains its minimum at $r_0$. Therefore $u'(r_0)=0$. Using these in (\ref{4psd2}),
we arrive at a contradiction to
the established fact $0<w(r_0)<1$. As a consequence of this fact and (\ref{4psd1}),
we see that $w'(r)<0$ for all $r>r_h$.
These derived properties will serve as major clues for our resolution of the
boundary value problem (\ref{4psd1}), (\ref{4psd2}), (\ref{t3}),
(\ref{t4})
which is to follow in the sequel.

Let us now consider the concrete case where $N$ is given as
\be\label{N1}
N(r)=1-\bigg(\frac{r_h}r\bigg)^{d-3},\quad r\geq r_h~,
\ee
which corresponds to a Schwarzschild background.
Our existence theorem for a nontrivial solution of (\ref{4psd1}), (\ref{4psd2}), (\ref{t3}),
(\ref{t4}) may be stated as follows.

\begin{theorem}\label{theoremt1} For the metric function $N$ defined by (\ref{N1}) and $\sigma=1$,
the boundary value problem (\ref{4psd1}), (\ref{4psd2}), (\ref{t3}),
(\ref{t4}) has a solution pair $(w,u)$ for some
constants $w_h\in (0,1]$ and
$\Phi>0$ so that both $w$ and $u$ are positive-valued functions
of the radial variable $r>r_h$ and $w$ strictly
increases.
\end{theorem}

In order to get a proof of the theorem, we shall again pursue a suitable
simplification of the system
of equations (\ref{4psd1}) and (\ref{4psd2}).
To this end, inserting (\ref{t5}) into  (\ref{4psd2}), we obtain
\be
[(1-w^2)^{p-1} N(\ln w)']'+\frac{(2p-1)}{r^2}(1-w^2)^p=0.\label{t6}
\ee
Next set $v=\ln w$ or $w=\e^v$. We can rewrite (\ref{t6}) as
\be
[(1-\e^{2v})^{p-1} N v']'+\frac{(2p-1)}{r^2}(1-\e^{2v})^p=0,\label{t7}
\ee
and arrive at the corresponding boundary condition
\be
v(r_h)=v_h=\ln w_h\leq 0,\quad v(\infty)=-\infty.\label{t8}
\ee
Moreover, using (\ref{t9}) and its inverse,
 we can again rewrite (\ref{t7}) into a semilinear equation,
\be
f''+f'(\ln N)'+\frac{(2p-1)}{r^2 N} (1-\e^{2Q(f)})^p=0,\quad r_h<r<\infty.\label{t10}
\ee

Set $r=\rho+r_h$. Then, in terms of the
differentiation with respect to $\rho>0$, we rewrite (\ref{t10}) as
\bea
\rho^2 f''&+&\rho f'\frac{(d-3)r_h^{d-3}\rho}{([\rho+r_h]^{d-3}-r_h^{d-3})(\rho+r_h)}\nn\\
&+&\frac{(2p-1)(\rho+r_h)^{d-5}\rho^2}{(\rho+r_h)^{d-3}-r_h^{d-3}}
(1-\e^{2Q(f)})^p=0,\quad 0<\rho<\infty.\label{t11}
\eea
With $t=\ln\rho$, we convert (\ref{t11}) into
\be
f_{tt}-f_t +g(t)f_t + h(t) (1-\e^{2Q(f)})^p=0,\quad -\infty<t<\infty,\label{t12}
\ee
subject to the boundary conditions
\be\label{t13}
f(-\infty)=-\alpha \quad (0\leq\alpha<\infty),\quad f(\infty)=-\infty,
\ee
where the functions $g(t)$ and $h(t)$ in (\ref{t12}) are defined by
\bea
g(t)&=&\frac{(d-3)r_h^{d-3}\e^t}{([\e^t+r_h]^{d-3}-r_h^{d-3})(\e^t+r_h)},\label{t14a}\\
h(t)&= &\frac{(2p-1)(\e^t+r_h)^{d-5}\e^{2t}}{(\e^t+r_h)^{d-3}-r_h^{d-3}},\label{t14b}
\eea
and $\alpha= -P(v_h)$ (see (\ref{t9})).

Recall that we are to solve (\ref{t12}) and (\ref{t13}) so that
its solution $f(t)$ is a negative-valued decreasing
function of $t$. For this purpose, we will use a shooting
method and consider the initial value problem
\bea
f_{tt}-f_t +g(t)f_t + h(t) (1-\e^{2Q(f)})^p&=&0,\quad -\infty<t<\infty,\label{t15}\\
f(t_0)&=&-a,\quad f_t(t_0)=-b,\quad\label{t16}
\eea
where $a, b>0$ and $t_0$ is fixed. Of course, consistency requires
\be\label{t17}
a>\alpha.
\ee

In order to realize the boundary condition $f(-\infty)=-\alpha$, we set $\tau=-t$,
$\tau_0=-t_0$, and convert (\ref{t15}) in the half interval
$-\infty<t\leq t_0$ into the form
\bea
f_{\tau\tau}+f_\tau -G(\tau)f_\tau&=&H(\tau)R(f),\quad \tau_0\leq\tau <\infty,\label{t18}\\
f(\tau_0)&=&-a,\quad f_\tau(\tau_0)=b,\quad\label{t19}
\eea
where $G(\tau)=g(-\tau)$ and $H(\tau)=h(-\tau)$ are both positive-valued and $R(\cdot)$ is defined by (\ref{tR})
as before.

For
fixed $a$ satisfying (\ref{t17}), we
use $f(\tau;b)$ to denote the unique solution of (\ref{t18}) and (\ref{t19}) which is defined in its
interval of existence.

To engage in a shooting analysis for (\ref{t18}) and (\ref{t19}), we define
\bea
{\cal B}^-&=&\{b\in\bfR\,|\,\exists \tau>\tau_0 \mbox{ so that } f_\tau (\tau;b)<0\},\nn\\
{\cal B}^0&=&\{b\in\bfR\,|\,f_\tau(\tau;b)>0\mbox{ and } f(\tau;b)\leq0\mbox{ for all }\tau>\tau_0\},\nn\\
{\cal B}^+&=&\{b\in\bfR\,|\,f_\tau(\tau;b)>0\mbox{ for all }\tau\geq\tau_0\mbox{ and }f(\tau;b)>0
\mbox{ for some }\tau>\tau_0\}.
\nn
\eea

\begin{lemma}\label{lemmat2}
We have the disjoint union $\bfR={\cal B}^-\cup{\cal B}^0\cup{\cal B}^+$.
\end{lemma}

\begin{proof} If $b\not\in{\cal B}^-$,
then $f_\tau(\tau;b)\geq0$ for all $\tau>\tau_0$. If there is a point $\tau_1>\tau_0$ so that
$f_\tau(\tau_1;b)=0$, then $f(\tau_1;b)\neq0$ because $f=0$ is an equilibrium point of the
differential equation (\ref{t18}) which is not attainable in finite $\tau$. Since
$f(\tau_1;b)\neq0$ but $f_\tau(\tau_1;b)=0$, we see that
either $f_{\tau\tau}>0$ or $f_{\tau\tau}<0$ at $\tau=\tau_1$. Hence, there is a $\tau<\tau_1$ or
$\tau>\tau_1$ at which $f_\tau(\tau;b)<0$ which implies $b\in {\cal B}^-$, a contradiction.
Thus, $f_\tau(\tau;b)>0$ for all $\tau>\tau_0$ and $b\in {\cal B}^0\cup{\cal B}^+$.
\end{proof}

\begin{lemma}\label{lemmat3}
The set ${\cal B}^-$ and ${\cal B}^+$ are both open and nonempty.
\end{lemma}

\begin{proof} The fact that ${\cal B}^-\neq\emptyset$ follows immediately from the
fact that $(-\infty,0)\subset {\cal B}^-$. The fact that ${\cal B}^-$ is open is self-evident.
To see that ${\cal B}^+$ is nonempty, first note that ${\cal B}^+\subset (0,\infty)$. Hence,
for $\tau>\tau_0$ but $\tau$ is close to $\tau_0$, we have $f_\tau>0$ and (\ref{t18}) gives us
\be\label{tt22}
(\e^\tau f_\tau)_\tau>\e^\tau H(\tau) R(f).
\ee
Integrating (\ref{tt22}) near $\tau_0$ where $f_\tau>0$, we have
\bea
f_\tau(\tau;b)&>&\bigg(b\e^{\tau_0}+\int_{\tau_0}^\tau H(s_1)R(f(s_1;b))\e^{s_1}\,\dd s_1\bigg)\e^{-\tau},\label{t20}\\
f(\tau;b)&>&-a+b(1-\e^{-(\tau-\tau_0)})\nn\\
&& +\int_{\tau_0}^\tau\e^{-s_2}\bigg(\int_{\tau_0}^{s_2}H(s_1)R(f(s_1;b))\e^{s_1}\,\dd
s_1\bigg)\,\dd s_2.\label{t21}
\eea
For any fixed $\tau_1>\tau_0$, we can choose $b>0$ sufficiently large so that
\bea
&&b\e^{\tau_0}+\int_{\tau_0}^{\tau_1} H(s_1) R(-a)\e^{s_1}\,\dd s_1>0,\label{t22}\\
&&-a+b(1-\e^{-(\tau_1-\tau_0)})+\int_{\tau_0}^{\tau_1}\e^{-s_2}\bigg(\int_{\tau_0}^{s_2}H(s_1)R(-a)\e^{s_1}\,\dd
s_1\bigg)\,\dd s_2>0.\label{t23}
\eea
In view of (\ref{t20})--(\ref{t23}), we see that there is a $\tau_2\in (\tau_0,\tau_1)$ so that
$f_\tau(\tau;b)>0$ for $\tau\in [\tau_0,\tau_2]$, $f(\tau;b)<0$ for $\tau\in[\tau_0,\tau_2)$,
but $f(\tau_2;b)=0$. Hence, for any $\tau>\tau_2$, there holds
\bea
f_\tau(\tau;b)&\geq&\bigg(b\e^{\tau_0}+\int_{\tau_0}^{\tau_2} H(s_1) R(f(s_1;b))\e^{s_1}\,\dd s_1\bigg)\e^{-\tau}\nn\\
&\geq& \bigg(b\e^{\tau_0}+\int_{\tau_0}^{\tau_1} H(s_1) R(-a)\e^{s_1}\,\dd s_1\bigg)\e^{-\tau}>0,\label{t24}\\
f(\tau;b)&>&0.\label{t25}
\eea
Therefore, $b\in {\cal B}^+$ and the nonemptyness of ${\cal B}^+$ is
established.

Moreover, for $b_0\in {\cal B}^+$, there is a $\tau_1>0$ so that $f(\tau_1;b_0)>0$. By the
continuous dependence of $f$ on the parameter $b$ we see that when $b_1$ is close to $b_0$
we have $f_\tau(\tau;b_1)>0$ for all $\tau\in[\tau_0,\tau_1]$ and $f(\tau_1,b_1)>0$. Using (\ref{t24})
again, we see that $f_\tau(\tau;b_1)>0$ for all $\tau>\tau_0$ as well, which proves
$b_1\in{\cal B}^+$. So ${\cal B}^+$ is open.

The fact that ${\cal B}^-$ is open is self-evident.
\end{proof}

\begin{lemma}\label{Lemma 2}
 The set ${\cal B}^0$ is nonempty and closed. Furthermore, if $b\in {\cal B}^0$, then $f(\tau;b)<0$ for all $\tau>\tau_0$.
\end{lemma}

\begin{proof} The first part of the lemma follows from the connectedness of $\bfR$, Lemma \ref{lemmat2}, and
Lemma \ref{lemmat3}.
To prove the second part, we assume otherwise that there is a $\tau_1>\tau_0$ so that
$f(\tau_1;b)=0$. Since $f(\tau;b)\leq0$ for all $\tau>\tau_0$, $f$ attains its local maximum at
$\tau_1$. In particular, $f_\tau(\tau_1;b)=0$, which contradicts the definition of ${\cal B}^0$.
\end{proof}

\begin{lemma}\label{Lemma 3}
 For $b\in {\cal B}^0$, there is a number $\alpha$ satisfying $0\leq\alpha<a$ such that $f(\tau;b)\to-\alpha$ as $\tau\to\infty$.
\end{lemma}

\begin{proof} Since $f$ increases as a function of $\tau\geq\tau_0$ and $f<0$ for all $\tau\geq\tau_0$, we see that the limit
$\lim_{\tau\to\infty}f(\tau;b)$
exists and satisfies $-a<lim_{\tau\to\infty} f(\tau;b)\leq0$.
\end{proof}

Returning to the original variable $t=-\tau$, we see that we have obtained a solution of (\ref{t12}) over the
left-half line $-\infty<t\leq t_0$ satisfying the boundary condition at $t=-\infty$ stated in (\ref{t13}).

We next
consider the problem over the right-half line $t_0\leq t<\infty$. For this purpose, let $f$ be a local solution
of (\ref{t15}) and (\ref{t16}) in a neighborhood of $t_0$. Since $1-g(t)>0$ and $h(t)>0$, we deduce from (\ref{t15})
that $f, f_t, f_{tt}$ all remain negative-valued for all $t\geq t_0$ in view of $f(t_0)<0$ and $f_t(t_0)<0$.
In particular, the solution is defined globally for all $t_0\leq t<\infty$.
Using
$h(\infty)=2p-1>0$, we see that $f_{tt}(t)\leq - c$ for some constant $c>0$ for all $t\geq t_0$. Consequently, we must
have $f(\infty)=-\infty$ which realizes the boundary condition at $t=\infty$ for $f$ stated in (\ref{t13}). In other
words, we have proved the existence of a solution of the two-point boundary value problem (\ref{t12}) and (\ref{t13}).

We are now ready to prove the existence theorem. To do so, we need only to examine the boundary conditions for the
original field functions $w$ and $u$ in terms of the radial variable $r$. Using the relations among various
variables, we obtain
\be
w_h\equiv \lim_{r\to r_h} w=\lim_{\rho\to 0} w=\lim_{\rho\to0}\e^v=\lim_{t\to-\infty} \e^{Q(f)}=\e^{Q(-\alpha)}\in (0,1],
\ee
as desired because $\alpha\geq0$ in view of Lemma \ref{Lemma 3}.

In order to realize the boundary condition for $u=-N(\ln w)'$ (see (\ref{t5})) at $r=r_h$, we insert the
definition of the function $N$ (see (\ref{N1})) to get
\bea\label{t30}
u&=&-N\frac{\dd v}{\dd r}=-\frac{([\rho+r_h]^{d-3}-r_h^{d-3})}{(\rho+r_h)^{d-3}}\frac{\dd v}{\dd \rho}\nn\\
&=&-\frac{(d-3)r_h^{d-4}(1+\mbox{O}(\rho))}{(\rho+r_h)^{d-3}}\rho\frac{\dd v}{\dd \rho}.
\eea
Using (\ref{t30}),  we obtain
\be\label{t31}
\lim_{r\to r_h} u=-(d-3)r_h^{-1}\lim_{t\to-\infty} \frac{\dd v}{\dd t}.
\ee

To evaluate the right-hand side of (\ref{t31}), recall that (\ref{t9}) gives us
\be\label{t32}
\frac{\dd v}{\dd t}=(1-\e^{2v})^{-(p-1)}\frac{\dd f}{\dd t}.
\ee

The easier case is when $w_h<1$. In view of (\ref{t31}) and (\ref{t32}), we have
\be\label{t33}
\lim_{r\to r_h} u=-(d-3)r_h^{-1}(1-w^2_h)^{-(p-1)}\lim_{t\to-\infty}\frac{\dd f}{\dd t}.
\ee
We again use the variable $\tau=-t$. Since $f$ satisfies
\be\label{t34}
f_{\tau\tau}+f_\tau(1-G(\tau))=-H(\tau) (1-\e^{2Q(f)})^p,\quad\tau\geq \tau_0,
\ee
we may integrate (\ref{t34}) to get
\be\label{t35}
f_\tau(\tau)=\e^{-\int_{\tau_0}^\tau(1-G(s))\,\tiny{\dd} s}\bigg(b-\int_{\tau_0}^\tau H(s)(1-\e^{2Q(f(s))})^p
\e^{\int_{\tau_0}^s(1-G(s_1))\,\tiny{\dd} s_1}\,\dd s\bigg).
\ee
The definitions of $G(s)$ and $H(s)$ give us the asymptotics
\bea\label{t36}
1-G(s)&=&\bigg(1+\frac{d-4}2\bigg)\e^{-s}+\mbox{O}(\e^{-2s}),\\
H(s)&=&\bigg(\frac{2p-1}{d-3}\bigg)r_h^{-1}\e^{-s}+\mbox{O}(\e^{-2s}),\label{t37}
\eea
for $s$ large. Using (\ref{t36}), (\ref{t37}), and the fact that $Q(f)<0$ in (\ref{t35}), we see that $f_\tau(\tau)$
is bounded for $\tau\geq\tau_0$. As a consequence, (\ref{t34}) leads us to the estimate
\be\label{t38}
f_{\tau\tau}(\tau)=\mbox{O}(\e^{-\tau})\quad\mbox{for large }\tau.
\ee
Since $f(\infty)=-\alpha$, we infer that there is a sequence $\{\tau_j\}$, $\tau_j\to\infty$ when $j\to\infty$, such that
$f_\tau(\tau_j)\to0$ when $j\to\infty$. From this fact and (\ref{t38}), we find
\be\label{t39}
f_\tau(\tau)=\mbox{O}(\e^{-\tau})\quad \mbox{for large }\tau.
\ee
Inserting (\ref{t39}) into (\ref{t33}), we arrive at the expected result
\be\label{t40}
\lim_{r\to r_h} u=0.
\ee

We now consider the case when $w_h=1$. Note that (\ref{t35})--(\ref{t39}) are all valid. Using
(\ref{t39}) and $f(\infty)=0$, we have
\be\label{t41}
f(\tau)=\mbox{O}(\e^{-\tau})\quad\mbox{for large }\tau.
\ee
Note also that (\ref{t9}) gives us the relation
\be\label{t42}
f=\int_0^v\bigg(-2s-\frac{(2s)^2}{2!}-\cdots\bigg)^{p-1}\,\dd s=(-2)^{p-1}\frac{v^p}{p} +\mbox{O}(v^{p+1}).
\ee
Combining (\ref{t41}) and (\ref{t42}), we obtain
\be\label{t43}
v(\tau)=\mbox{O}(\e^{-\tau/p})\quad \mbox{when $\tau$ is large}.
\ee
Inserting (\ref{t39}) and (\ref{t43}) into (\ref{t32}), we get
\be\label{t44}
\frac{\dd v}{\dd t}=v^{-(p-1)}(-2+\mbox{O}(v))^{-(p-1)} \frac{\dd f}{\dd t}=\mbox{O}(\e^{t/p})\quad\mbox{ as }t\to-\infty,
\ee
which establishes (\ref{t40}) again in view of (\ref{t31}).

We finally examine the behavior of the solution pair $(w,u)$ at $r=\infty$. Since we have derived the limit
$f(t)\to-\infty$ as $t\to\infty$, we have
\be\label{t45}
\lim_{r\to\infty}w=\lim_{\rho\to\infty} \e^v=\lim_{t\to\infty} \e^{Q(f)}=0.
\ee
Besides, using the definition (\ref{N1})
for the background function $N$, the relation $u=-N(\ln w)_r$, and $v(\infty)=-\infty$,
we have
\be\label{t46}
\lim_{r\to\infty} u=-\lim_{\rho\to\infty}\frac{\dd}{\dd\rho}(\ln w)=-\lim_{t\to\infty}\e^{-t}\frac{\dd v}{\dd t}
=-\lim_{t\to\infty}\e^{-t}\frac{\dd f}{\dd t}.
\ee
On the other hand, multiplying (\ref{t12}) by
$
\e^{-\int_{t_0}^t (1-g(s))\,\tiny{\mbox{d}} s}
$
and integrating over $(t_0,t)$, we get
\be\label{t47}
f_t(t)=\e^{\int_{t_0}^t (1-g(s))\,\tiny{\mbox{d}} s}\bigg(-b-\int_{t_0}^t h(s)(1-\e^{2Q(f(s))})^p\e^{-\int_{t_0}^s (1-g(s_1))\,\tiny{\mbox{d}} s_1}\,\dd s\bigg),
\ee
which results in the expression
\bea\label{t48}
-\lim_{t\to\infty}\e^{-t}f_t(t)
&=&\e^{-t_0-\int_{t_0}^\infty g(s)\,\tiny{\mbox{d}} s}\bigg(b+\int_{t_0}^\infty h(s)(1-\e^{2Q(f(s))})^p\e^{-\int_{t_0}^s (1-g(s_1))\,\tiny{\mbox{d}} s_1}\,\dd s\bigg)\nn\\
&\equiv&\Phi>0,
\eea
where we have used the properties
\be\label{t49}
g(t)=\mbox{O}(\e^{-(d-3)t}),\quad h(t)=\mbox{O}(1)\quad\mbox{ as }t\to\infty,
\ee
for the positive-valued functions $g(t)$ and $h(t)$ given in (\ref{t14a}) and (\ref{t14b}) to deduce the convergence of the two
improper integrals in (\ref{t48}). In other words, in view of (\ref{t46}), the positive number $\Phi$ defined in (\ref{t48}) gives us the expected limit,
\be\label{t50}
\lim_{r\to\infty} u =\Phi.
\ee

The proof of the existence theorem is now complete.
%\medskip
%%%%%%%%%%%%%%%%%%%%%%%%%%%%%%%%%%%%%%%%%%%%%%%%%%%%%%%%%%%%
\subsubsection{The proof for a general black hole background}
%%%%%%%%%%%%%%%%%%%%%%%%%%%%%%%%%%%%%%%%%%%%%%%%%%%%%%%%%%%%
We now turn our attention to the case where the blackhole metric is defined by
general functions, $N>0$ and $\sigma>0$
(whenever $r>r_h$).
Focusing again on the selfdual equations with the choice of upper signs,
we have (\ref{4psd1}) and (\ref{4psd2})
subject to the same boundary conditions, (\ref{t3}) and (\ref{t4}).
Note that we are interested again in nonnegative-valued
solutions.  As discussed earlier,  when $w_h=0$, the solution is unique and explicitly given by
\be
w\equiv0,\quad u(r)=(2p-1)\int_{r_h}^r\frac{\sigma(\rho)}{\rho^2}\,\dd\rho,\quad \Phi=(2p-1)\int_{r_h}^\infty\frac{\sigma(\rho)}{\rho^2}\,\dd\rho,
\ee
which may be called a trivial solution; when $w_h\in(0,1]$, however, the solution will not be trivial. Furthermore, we can
see that if $(w,u)$ is a solution with $w_h>0$, then $w(r)>0, u(r)>0$ for $r>r_h$ and $w$
strictly increases. In particular, (\ref{4psd1}) allows us to represent $u$ by (\ref{t1.7})
which of course
generalizes (\ref{t5}). Therefore, using the same substitution of variables, $v=\ln w$ and $f=P(v)$ as given in
(\ref{t9}), we can transform (\ref{4psd1}) and (\ref{4psd2}) into the scalar equation (\ref{t54}).

Like before, we write $r=\rho+r_h$ $(\rho>0)$. Therefore, in terms of $t=\ln\rho$, we convert
(\ref{t54}) into the familiar form (\ref{t12}) in which the coefficient functions $g(t)$ and $f(t)$ are now defined by
the updated expressions
\bea\label{t55a}
g(t)&=&\bigg[\rho\bigg(\frac{\sigma'(r)}{\sigma (r)}+\frac{N'(r)}{N(r)}\bigg)\bigg]_{\rho=\e^t},\quad \\
h(t)&=&\bigg[(2p-1)\frac{\rho^2}{r^2N(r)}\bigg]_{\rho=\e^t}.\label{t55b}
\eea

Recall that the key properties of the functions $g(t)$ and $h(t)$ used in the proof of Theorem \ref{theoremt1} are

(i) $g(t)\geq 0$ and $h(t)>0$ for all $t$;

(ii) $1-g(t)>0$ for all $t$;

(iii) $h(\infty)>0$;

(iv) as $t\to-\infty$, we have the asymptotics
\be\label{t56}
1-g(t)=\mbox{O}(\e^{\varepsilon t}),\quad h(t)=\mbox{O}(\e^{\delta t}),
\ee
for some $\varepsilon, \delta>0$;

(v) as $t\to\infty$, we have the asymptotics
\be\label{t57}
g(t)=\mbox{O}(\e^{-\gamma t}),\quad h(t)=\mbox{O}(1),
\ee
for some $\gamma >0$.

It is direct to check that the proof of Theorem \ref{theoremt1} is intact for the general $N,\sigma$ case
when the above properties are valid. Consequently, switching back to the original radial variable $r$, we arrive
at the following general existence theorem.

\begin{theorem}\label{theoremt2}
For the general system of BPS equations (\ref{4psd1}) and (\ref{4psd2})
subject to the boundary condition given by (\ref{t3}) and (\ref{t4}), the same conclusion for existence of a solution
as stated in Theorem \ref{theoremt1} holds provided that the
positive-valued metric functions $N$ and $\sigma$ satisfy the
assumptions
\be\label{t58}
0\leq (r-r_h)\bigg(\frac{\sigma'(r)}{\sigma(r)}+\frac{N'(r)}{N(r)}\bigg)<1,\quad \forall r>r_h,
\ee
\be\label{t59}
\lim_{r\to\infty} N(r)=N_\infty>0,
\ee
\be \label{t60}
1-(r-r_h)\bigg(\frac{\sigma'(r)}{\sigma(r)}+\frac{N'(r)}{N(r)}\bigg)=
\mbox{O}((r-r_h)^\varepsilon), \quad r\approx r_h,
\ee
\be \label{t61}
\frac{(r-r_h)^2}{N(r)}=\mbox{O}((r-r_h)^\delta),\quad r\approx r_h,
\ee
\be \label{t62}
(r-r_h)\bigg(\frac{\sigma'(r)}{\sigma(r)}+\frac{N'(r)}{N(r)}\bigg)=
\mbox{O}((r-r_h)^{-\gamma}),\quad r\approx\infty,
\ee
where $\varepsilon, \delta, \gamma$ are some positive constants.
\end{theorem}

The remarks at the end of the subsection 4.1.2 on the generality of
these conditions apply in this case, too. Again, the only condition
which seems to require a knowledge of $N,\sigma$ is (\ref{t58}).
However, (\ref{t58}) is satisfied by the various backgrounds we have
considered.
%
%%%%%%%%%%%%%%%%%%%%%%%%%%%%%%%%%%%%%%%%%%%%%%%%%%%%%%%%%%%%
\subsubsection{The proof for a topologically trivial background}
%%%%%%%%%%%%%%%%%%%%%%%%%%%%%%%%%%%%%%%%%%%%%%%%%%%%%%%%%%%%
We now consider type II YM instantons on a `soliton' background for which the
functions $N$ and $\sigma$ are
everywhere positive and regular up to $r=0$:
\be \label{t63}
N(0)=1,\quad \sigma(0)=\sigma_0\quad\mbox{with } 0<\sigma_0<1.
\ee
The governing equations are still (\ref{4psd1}) and (\ref{4psd2}) subject to the boundary conditions (\ref{t3}) and (\ref{t4}).
In this situation, we may simply restate our sufficient conditions given in Theorem \ref{theoremt2} by setting
$r_h=0$ in order to guarantee the existence of a solution with the only exception that (\ref{t60}) is
no longer valid when $r_h=0$
and the metric functions $N$ and $\sigma$ are regular at $r=0$ which gives rise to the new property
\be \label{t64}
\lim_{r\to0}r\bigg(\frac{\sigma'(r)}{\sigma(r)}+\frac{N'(r)}{N(r)}\bigg)=0.
\ee
In fact, we note that, with $r_h=0$, $r=\rho=\e^t$, and
\bea\label{t65a}
g(t)&=&\bigg[r\bigg(\frac{\sigma'(r)}{\sigma(r)}+\frac{N'(r)}{N(r)}\bigg)\bigg]_{r=\e^t},\quad\\
h(t)&=&\bigg[\frac{(2p-1)}{N(r)}\bigg]_{r=\e^t},\label{t65b}
\eea
the equations (\ref{4psd1}) and (\ref{4psd2}) are condensed into the equation (\ref{t12}) as before. Thus, in view
of (\ref{t63})--(\ref{t65b}), we have
\be\label{t66}
\lim_{t\to-\infty} g(t)=0,\quad \lim_{t\to-\infty}h(t)=2p-1,
\ee
which violates the condition (\ref{t56}). Therefore, our existence theorem may not be applicable directly to this
`regularised' problem. Below we shall show that (\ref{t66}) actually allows us to strengthen our existence
theorem.

First, it is straightforward to see that (\ref{t66}) renders no barrier to the existence of a solution of
(\ref{t12}) subject to the boundary condition (\ref{t13}) which may be obtained as before through solving
the initial value problem (\ref{t15})--(\ref{t16}) for given $t_0$, $a>0$, and a suitable $b>0$ (recall that
the set ${\cal B}^0$ contains all such suitable $b$'s).

\begin{lemma}\label{lemmat7}
Let $f$ be the solution of (\ref{t15})--(\ref{t16}) with $b\in {\cal B}^0$. Then $\alpha=0$ or $f(-\infty)=0$.
In particular, ${\cal B}^0$ contains exactly one point $b=b(a,t_0)$ (say) and
\be\label{t67}
{\cal B}^-=(-\infty, b(a,t_0)),\quad {\cal B}^+=(b(a,t_0),\infty).
\ee
\end{lemma}
\begin{proof}
Use the variable $t=-\tau$ and consider instead the problem (\ref{t18})--(\ref{t19}) for $\tau\geq \tau_0$. We then
arrive at (\ref{t35}) with $b\in{\cal B}^0$. Recall that (\ref{t66}) implies that
\be\label{t68}
\lim_{s\to\infty} G(s)=0,\quad \lim_{s\to\infty} H(s)=2p-1.
\ee
If $\alpha>0$, then $f(\tau)<-\alpha$ for all $\tau\geq\tau_0$ and we conclude from (\ref{t35}) that $f_\tau(\tau)<0$ when $\tau$ is sufficiently large, which
contradicts the definition of ${\cal B}^0$.

If $b_1, b_2\in {\cal B}^0$, then $z(\tau)=f(\tau;b_1)-f(\tau;b_2)$ satisfies the boundary condition $z(\tau_0)=
z(\infty)=0$. Inserting this into (\ref{t18}), we have
\be\label{t69}
z_{\tau\tau}+(1-G(\tau))z_\tau= H(\tau) R'(\xi)z,\quad \tau\geq \tau_0,
\ee
where $\xi$ lies between $f(\tau; b_1)$ and $f(\tau;b_2)$. Applying the fact that $H(\tau)R'(\xi)>0$ and the maximum
principle in (\ref{t69}), we get $z\equiv0$. In particular, $b_1=b_2$. So ${\cal B}^0$ contains exactly one point
as claimed.
\end{proof}

To simplify notation, we set $t_0=0$ in (\ref{t48}). For any $a>0$ in the initial value problem
(\ref{t15})--(\ref{t16}), let $b=b(a)$ be the unique point in ${\cal B}^0$ ensured by the above
lemma. Hence we can rewrite the boundary value $\Phi$ given in
(\ref{t48}) as a well-defined function of $a$, say $\Phi(a)$, as follows,
\be\label{t70}
\Phi(a)=\e^{-\int_{0}^\infty g(s)\,\tiny{\mbox{d}} s}\bigg(b(a)+\int_{0}^\infty h(s)(1-\e^{2Q(f(s))})^p\e^{-\int_{0}^s (1-g(s_1))\,\tiny{\mbox{d}} s_1}\,\dd s\bigg).
\ee

\begin{lemma}\label{lemmat8}
The function $\Phi(\cdot)$ depends on $a>0$ continuously
 so that $\Phi(a)\to0^+$ as $a\to0^+$ and $\Phi(a)\to\infty$
as $a\to\infty$. In particular, the range of $\Phi(\cdot)$ is the entire half interval
$(0,\infty)$.
\end{lemma}

\begin{proof} First, we show that $b(a)$ is continuous with respect to the parameter $a>0$.
Otherwise there is a point $a_0>0$ and a sequence $\{a_j\}\subset(0,\infty)$ so that $a_j\to a_0$
as $j\to\infty$ but
$|b(a_j)-b(a_0)|\geq\vep_0$ for some $\vep_0>0$ and $j=1,2,\cdots$. From the proof of Lemma \ref{lemmat3} (cf. (\ref{t20})--(\ref{t23})), we see that $\{b(a_j)\}$ is a bounded sequence. In fact, with $\tau_0=0$ and $\tau_1=1$ in
(\ref{t20})--(\ref{t23}), we see that $b\in {\cal B}^+$ when (\ref{t22}) and (\ref{t23}) or
\be\label{t71}
b>\Gamma_1(a)\equiv\int_0^1 H(s)(1-\e^{2Q(-a)})^p\e^s\,\dd s
\ee
and
\be
b(1-\e^{-1})>\Gamma_2(a)\equiv a+\int_0^1 \e^{-s_2}\bigg(\int_0^{s_2} H(s_1)(1-\e^{2Q(-a)})^p\e^{s_1}\,\dd s_1\bigg)\,\dd s_2\label{t72}
\ee
hold. In other words, (\ref{t71}) and (\ref{t72}) give us the upper bound
\be\label{t73}
b(a)\leq \max\{\Gamma_1(a),(1-\e^{-1})^{-1} \Gamma_2(a)\}.
\ee
In particular, the boundedness of $\{b(a_j)\}$ follows.
Hence, passing to a subsequence if necessary, we may assume $b(a_j)\to$ some $b_0$ as $j\to\infty$. Of course,
$b_0\neq b(a_0)$.
It is clear that for (\ref{t18})--(\ref{t19}) with $a=a_0$, both $b(a)$ and $b_0$ lie in ${\cal B}^0$, which contradicts
Lemma \ref{lemmat7} which states that ${\cal B}^0$ contains exactly one point.

The continuous dependence of $b(a)$ on $a$ implies that the solution $f$ of (\ref{t18})--(\ref{t19}) with $b=b(a)$
depends on $a$ continuously as well. Using this
fact, we easily obtain the continuous dependence of $\Phi(a)$ on $a>0$
because the improper integral containing $f$ on the right-hand side of (\ref{t70})
is uniformly convergent with respect to the parameter $a>0$ in view of (\ref{t57}).

We claim that $b(a)\to 0^+$ as $a\to 0^+$. Otherwise there is a sequence $\{a_j\}$ in $(0,\infty)$ and
an $\vep_0$ so that $a_j\to0$
as $j\to\infty$ but $b(a_j)\geq\vep_0$ ($j=1,2,\cdots$). Using these in the initial value problem
 (\ref{t18})--(\ref{t19})  with
$a=a_j$ and $b=b(a_j)$, we observe that the solution will assume a positive value for a slightly positive $\tau$
when $j$ is sufficiently large, which
contradicts the definition of $b(a_j)$.

We can also claim that $b(a)\to\infty$ as $a\to\infty$. To see this, we insert the property $f_\tau\geq0$ in
(\ref{t35}) (with $\tau_0=0$) to get
\be \label{t74}
b(a)\geq\int_0^\tau H(s)(1-\e^{2Q(f(s))})^p \e^{\int_0^s (1-G(s_1))\tiny{\dd} s_1}\,\dd s,\quad \forall \tau\geq0.
\ee
Since we also know that $f_{\tau\tau}\leq0$ for $\tau\geq0$, we have
\be\label{t75}
f(\tau)\leq -a+b(a)\tau \quad\mbox{for }\tau\geq0.
\ee
In view of (\ref{t75}), we see that $f(\tau)\leq -1$ (say) whenever $\tau$ satisfies
\be\label{t76}
\tau\leq\frac{(a-1)}{b(a)},\quad a>1.
\ee
Combining (\ref{t74}) and (\ref{t76}), we get the lower bound
\be \label{t77}
b(a)\geq\int_0^{(a-1)/b(a)} H(s)(1-\e^{2Q(-1)})^p \e^{\int_0^s (1-G(s_1))\tiny{\dd} s_1}\,\dd s,
\ee
which implies that $b(a)\to\infty$ as $a\to\infty$ because the integral
\be
\int_0^{\infty} H(s)(1-\e^{2Q(-1)})^p \e^{\int_0^s (1-G(s_1))\tiny{\dd} s_1}\,\dd s
\ee
is divergent which allows us to argue by contradiction if $b(a)\not\to \infty$ when $a\to\infty$.

Hence $\Phi(a)$ defined in (\ref{t70}) is a continuous function with respect to $a\in(0,\infty)$
so that $\Phi(a)\to 0^+$ when $a\to 0^+$ and $\Phi(a)\to\infty$ when $a\to\infty$.

The proof of the lemma is now complete.
\end{proof}

In summary, we can state our results for the existence of a solution in the regular case as follows.

\begin{theorem}\label{theoremt3}
Suppose that the background metric functions $N$ and $\sigma$ are regularly defined for all $r>0$ and satisfy
(\ref{t63}) at $r=0$ and the conditions (\ref{t58}), (\ref{t59}), (\ref{t61}),  (\ref{t62})
(all with $r_h=0$), and (\ref{t64}).  Then the system of equations (\ref{4psd1})
and (\ref{4psd2}) subject
to the boundary conditions (\ref{t3}) and (\ref{t4}) governing a BPS monopole has a nontrivial solution if and
only if $w_h=1$. Moreover, the positive constant $\Phi$ in (\ref{t4}) may be taken to be any prescribed number
and for any given $\Phi>0$, the solution pair $(w,u)$ satisfies $w>0, u>0$, and $w$ strictly increases for $r>0$.
\end{theorem}

%%%%%%%%%%%%%%%%%%%%%%%%%%%%%%%%%%%%%%%%%%%%%%
\section{Solutions with spherical symmetry \\ in (even) $d$-dimensions }
\setcounter{equation}{0}
%%%%%%%%%%%%%%%%%%%%%%%%%%%%%%%%%%%%%%%%%%%%%

Here we start by imposing spherical symmetry in the whole $d$ dimensional
 (Euclidean) space, treating all coordinates on the same footing.
The corresponding metric Ansatz in this case is
\be
\label{metric2}
ds^2=d\rho^2+f^2(\rho)d \Omega_{(d-1)}^2,
\ee
where $f(\rho)$ is a function fixed by the gravity-matter field equations,
$\rho$ being the radial coordinate (with $\rho_a\leq \rho \leq \rho_b$).
%($e.g.$ $f(\rho)=\rho$ for flat space, $f(\rho)=\rho_0 \sinh (\rho/\rho_0)$ for an
%AdS background etc).

The YM Ansatz compatible with the symmetries of the above line element is
expressed as
\be
\label{YMsph2}
A_{\mu}=\left(\frac{1-w(\rho)}{\rho}\right)\,\Sigma_{\mu\nu}^{(\pm)}\hat x_{\nu}\,,
\ee
where the spin matrices are precisely those used in \re{YMhier}, the
radial variable in
is $\rho_{\mu}=\sqrt{|x_{\mu}|^2}$ and $\hat x_{\mu}=x_{\mu}/\rho$ is
the unit radius vector.

The resulting reduced one dimensional YM Lagrangian for the $p$-th term in the YM ierarchy
is
\bea
L^{(p,d)}_{\rm{YM}}&=&\frac{\tau_p}{2\cdot (2p)!}
\frac{(d-1)!}{(d-2p)!}
f^{d-4p+1} (w^2-1)^{2p-2}\left(w'^2+\frac{d-2p}{2p}\frac{(w^2-1)^2}{f^2}\right)~,
\label{LYMpd-s2}
\eea
the corresponding $d=4p$ YM selfduality equations taking the simple form
\bea
w'\pm \frac{w^2-1}{f(\rho)}=0.
\label{eq-s2}
\eea
For any choice of the metric function $f(\rho)$, the solution of the
above equation reads
\bea
w=\frac{1-c_0e^{\mp 2\int \frac{d\rho}{f(\rho)}}}{1+c_0e^{\mp 2\int \frac{d\rho}{f(\rho)}}}
\label{eq-s3}
\eea
where $c_0$ is an arbitrary positive constant.

The action of the selfdual solutions can be written as
\bea
S=\pm 2\frac{\tau_p(4p-1)!}{2\cdot ((2p)!)^2}
V_{d-1}w_{~2}F_1(\frac{1}{2},1-2p,\frac{3}{2},w^2)\bigg|_{\rho_0}^{\rho_1}
\eea
($_2F_1(a,b,c,z)$ being the hypergeometric function).
For $f(\rho)=\rho$ one recovers the $d=4p$ generalization of the
BPST instanton first found in \cite{Tchrakian:1984gq}, with $w=(\rho^2-c)/(\rho^2+c)$.
An AdS background $f(\rho)=\rho_0\sinh \rho/\rho_0$ leads to a
$d=4p$ generalization of the $d=4$ AdS selfdual instantons in \cite{Maldacena:2004rf}, with
$w=(\tanh^2(\rho/2\rho_0)-c)/(\tanh^2(\rho/2\rho_0)+c)$.
The $d=4p$ selfdual instantons on a sphere (euclideanised dS space)
are found by taking $\rho_0\to i \rho_0$ in the corresponding AdS relations.

Again, one can consider as well the superposition of two members of the
YM hierarchy, say those labeled by $p$ and $q$, with $d=2(p+q)$.
The generic selfduality equations (\ref{sdpq}) reduce here to
\bea
\label{sdpqred}
w'\pm \sqrt{\frac{\kappa p}{q}}(1-w^2)^{q-p+1}f^{2p-2q-1}=0,~~
w'\pm \sqrt{\frac{q}{\kappa p}}(1-w^2)^{p-q+1}f^{2q-2p-1}=0~,
\eea
where we note $\kappa=\hat \tau_p/\hat \tau_q$,
which is supposed to be a positive quantity.
The above are overdetermined and have
the unique solutions
\be
\label{kihara}
f=r_0\sin \rho/\rho_0,~w=\cos \rho/\rho_0\ \ {\rm or}\ \
f=r_0\sinh \rho/\rho_0,~w=\cosh \rho/\rho_0,\ \ {\rm with}\ \
r_0=(\kappa p/q)^{\frac{1}{4(q-p)}}\,,
\ee
accommodating fixed AdS and dS spaces, depending if $\ka$ is negative or positive.
Note that when $\ka$ is
negative, $r_0$ in \re{kihara} cannot be real,
so we have really only a YM field on dS.
Here, unlike in the $d=4p$ case featuring the single $F(2p)$,
the expression of the metric function $f(\rho)$
is fixed to describe a curved
maximally symmetric background.
The dS case with $p=1,~q=2$ solution of \re{kihara} was recently
found in \cite{Kihara:2007di}.

%%%%%%%%%%%%%%%%%%%%%%%%%%%%%%%%%%%%%%%%%%%%%%
\section{Summary and discussion}
\setcounter{equation}{0}
%%%%%%%%%%%%%%%%%%%%%%%%%%%%%%%%%%%%%%%%%%%%%%
We have considered the problem of constructing instantons of gravitating Yang--Mills
field systems in all even
dimensions. Our constructions are limited to two highly symmetric kinds of solutions
for which the effective
field equations are one dimensional. The larger part of the work concerns (Euclidean time)
static fields
that are spherically symmetric in the $d-1$ space dimensions, and a smaller
 part deals with fields that are
spherically symmetric in the whole $d$ dimensional (Euclidean) spacetime,
treating all coordinates on the same
footing. The main task here was the extension of the known
results~\cite{Charap:1977ww,Brihaye:2006bk,Maldacena:2004rf} in $d=4$ to arbitrary even $d$.

The static, and spherically symmetric (in $d-1$ dimensions) solutions are
interesting because in that case
instantons on (Euclideanised) black holes can be described. The fully spherically symmetric
(in $d$ dimensions) solutions
are also interesting because in that case the metric is a conformal deformation
of the metric on $S^{4p}$,
relevant to the $p-$BPST instantons. All cases studied are restricted to
Yang--Mills selfdual solutions,
which means that the gravitational background is a fixed one offering no
backreaction to the YM field,
since the stress tensor of selfdual YM fields vanishes. We have nontheless
searched numerically
for solutions to the second order field equations that might describe radial excitations of the
selfdual ones we construct, and have found no such solutions.

The static spherically symmetric solutions we have studied come in two types. Type I are the $d$
dimensional generalisations of the Charap--Duff~\cite{Charap:1977ww} (CD) solution in $4$ dimensions,
while Type II generalise the deformed Prasad--Sommerfield
(PS) monopole in~\cite{Brihaye:2006bk}.
For $d=4p$ we give the exact Type I analogues of the CD solution
on double-selfdual $p-$Schwarzschild
backgrounds, including a cosmological constant, in closed form.
On backgrounds that are not double-selfdual,
the solutions are constructed numerically.
 In our numerical constructions we have mostly employed backgrounds
arising from $p-$Einstein gravity, notably the $p-$Reissner--Nordstr\"om.
In addition, we have verified that
$p-$YM instantons of Type I satisfying CD like boundary conditions can
be constructed numerically on
$q-$Einstein backgrounds ($p\neq q$), but these are much less robust that when $p=q$.
In $d=4p+2$ the only
Type I selfdual solutions are those
on fixed AdS$_{2p}$ and dS$_{2p}$ backgrounds, evaluated in closed form,
and, in the AdS case the solution is not real.
Type II solutions are evaluated only numerically, and
only in $4p$ dimensions. These are deformations of $p-$Prasad-Sommerfield
monopoles. Both Type I an II solutions describe the YM field on a fixed black hole,
 but while the radius
of the horizon $r_h$ for Type I is unconstrained, $r_h$ for a Type
II solution has a maximal value. The numerical results presented in
section 3 are supported by the existence proofs given in section 4.
In evaluating the numerical instantons of both Types, we solved the
second order field equations, but found no radial excitations above
the selfdual solutions. The existence of all numerically constructed
solutions was proved analytically using a dynamic shooting method.

The last type of instantons considered in this paper, namley those deforming the $p-$BPST instanton
on $\R^{4p}$, are evaluated in closed form both on AdS and dS backgrounds. In $4p+2$ dimensions, the
selfduality equations yield the same fixed backgrounds as in the case of Type I solutions in these dimensions.

Perhaps one of the most remarkable qualitative features of Types I and II instantons is, that these are not
instantons at all but rather are monopole like lumps. This is because for a
genuine instanton, the radial function
$w$ appearing in the Ansatz \re{YMsph} must change sign over the full range $r_h$ to infinity of the radial
coordinate, while in what we have here, the sign of $w$ does not
change in this range. This behaviour is typical of a monopole. This aspect of
the CD solutions is consistent
with the conclusion of Tekin~\cite{Tekin:2002mt}, who has allowed a time dependent YM field on
the $1-$Schwarzschild background in $4$ dimensions, and found that the resulting solution
remains static,
namely the CD solution itself. This conclusion is clearly true also in the $4p$ dimensional $p-$CD instantons
here. We intend to carry this line of investigation further.
%It would be interesting to further study their connection with the caloron physics.

\bigskip

\noindent {\bf Acknowledgements}
\\
D.H.T. is deeply indepted to A. Chakrabarti for past collaboration
on this subject.
The work of E.R. and D.H.T. was carried out in the framework of
Science Foundation Ireland (SFI) Research Frontiers Programme (RFP)
project RFP07/FPHY330. The research of Y.Y. was supported in part by
National Science Foundation under grant DMS--0406446 and an Othmer
Senior Faculty Fellowship at Polytechnic University.

%%%%%%%%%%%%%%%%%%%%%%%%%%%%%%%%%%%%%%%%%%%%%
\appendix
\section{Double selfdual spaces}
\setcounter{equation}{0}
\renewcommand{\theequation}{A.\arabic{equation}}
The considered gravitational system in $d=2(p+q)$ spacetime dimensions
is the superposition of all possible scalars $R_{(p,q)}$
\be
\label{EHhier}
{\cal L}_{\rm{grav}}\ =\ \sum_{p=1}^{P}\
\frac{\ka_{p}}{2p}\ e \ R_{(p,q)}\ ,
\ee
where $R_{(p,q)}$ are constructed from the $2p$-form $R(2p)=R\wedge R\wedge
...\wedge R$ resulting from the totally antisymmetrised $p$-fold
products of the Riemann curvature $2$-forms $R$. We express
$R_{(p,q)}$ in the notation of \cite{O'Brien:1988rs} as
\be
\label{gp}
e\,R_{(p,q)}=\vep^{\mu_1\mu_2...\mu_{2p}\nu_1\nu_2...\nu_{2q}}
e_{\nu_1}^{n_1}e_{\nu_2}^{n_{2q}}...e_{\nu_{2q}}^{n_{2q}}
~\vep_{m_1m_2...m_{2p}n_1n_2...n_{2q}}\,
R_{\mu_1\mu_2....\mu_{2p}}^{m_1m_2..m_{2p}}\,,
\ee
where $e_{\nu}^{n}$ are the {\it Vielbein} fields,
$e=\mbox{det}(e_{\nu}^{n})$ in \re{EHhier}, and
$R_{\mu_1\mu_2....\mu_{2p}}^{m_1m_2..m_{2p}}=R(2p)$ is the $p$-fold totally
antisymmetrised product of the Riemann curvature, in component notation. It is
clear from the definition \re{gp} that the spacetime dimensionality is $d=2(p+q)$,
and that the maximum value $p$ in the sum \re{EHhier} is $P=\frac12(d-2)$, with
the term $e\,R_{(p=\frac{d}{2},q=0)}$ being the (total divergence)
Euler-Hirzebruch density. Subjecting \re{gp} to the variational principle one
arrives at the $p$-th order Einstein equation
\be
\label{peinstein}
G_{(p)}{}_{\mu}^m=R_{(p)}{}_{\mu}^m\
-\ \frac{1}{2p}\ R_{(p)}\ e_{\mu}^m\,,
\ee
in terms of the $p$-th order Einstein tensor $G_{(p)}{}_{\mu}^m$,
with $R_{(p)}$ and $R_{(p)}{}_{\mu}^{m}$ being the $p$-th order Ricci scalar and
the $p$-th order Ricci tensor defined respectively by
\bea
R_{(p)}&=&R_{\mu_1\,\mu_2....\mu_{2p}}^{m_1\,m_2..m_{2p}}\;
e^{\mu_1}_{m_1}\,e^{\mu_1}_{m_2}\,...\,e^{\mu_{2p}}_{m_{2p}}\label{ricci-s}\\
R_{(p)}{}_{\mu}^{m}&=&R_{\mu\,\mu_2....\mu_{2p}}^{m\,m_2..m_{2p}}\;
e^{\mu_2}_{m_2}\,...\,e^{\mu_{2p}}_{m_{2p}}\;\;.\label{ricci-t}
\eea

Let us first consider the special case of $p=q$, namely of a $4p$ dimensional
spacetime. The double selfduality condition in that case is
\be
\label{dsd}
R_{\mu_1\mu_2....\mu_{2p}}^{m_1m_2..m_{2p}}=\pm
\frac{e}{[(2p)!]^2}\,
\vep_{\mu_1\mu_2...\mu_{2p}\nu_1\nu_2...\nu_{2p}}
R^{\nu_1\nu_2....\nu_{2p}}_{n_1n_2..n_{2p}}\,
~\vep^{m_1m_2...m_{2p}n_1n_2...n_{2p}}\,,
\ee
The $\pm$ sign in \re{dsd} pertains to Euclidean and Minkowskian
signatures, respectively, which is in order to impose in this case
unlike in the case of single selfduality in which case the Hodge
dual for Minkowskian signature would introduce an undesirable factor
of $i=\sqrt{-1}$. We shall soon see that it is gainful to keep only
to Euclidean signature.

Contracting the l.h.s. of \re{dsd} with
$e^{\mu_2}_{m_2}\,e^{\mu_2}_{m_2}\,...\,e^{\mu_{2p}}_{m_{2p}}$, and
relabeling the free indices $(\mu_1,m_1)$ as $(\mu,m)$ we get the $p$-th
order Ricci tensor defined by \re{ricci-t}. After applying the usual
tensor identities this results in the constraint
\be
\label{constrp}
R_{(p)}{}_{\mu}^m\ =\ \mp\left(R_{(p)}{}_{\mu}^m\
-\frac{1}{2p}\,R_{(p)}\,e_{\mu}^m\right)\,.
\ee
It is now obvious that in the case of Minkowskian signature (the lower sign)
\re{constrp} leads to
\be
\label{weakp}
R_{(p)}=0
\ee
which is too weak a constraint to satisfy any $p-$Einstein equation arising from the
variation with respect to $e_m^{\mu}$. Accordingly, we restrict to Euclidean
spaces, whence \re{constrp} reads
\be
\label{dsd-p}
G_{(p)}{}_{\mu}^m\ =\ -\frac{1}{4p}\,R_{(p)}\,e_{\mu}^m\,.
\ee
The most general Lagrangian whose field equations are solved by the constraint
\re{dsd-p} is the following special case of \re{EHhier} augmented with a cosmological
constant $\Lambda$,
\be
\label{EHp}
{\cal L}_{\rm{grav}}\ =
\ e\ \left(\ka^{2p}\,R_{(p,p)}\ +\ d!\ \Lambda\right)\,,
\ee
where $\ka$ is a constant with the dimension of a length. The Einstein equations of
\re{EHp} are
\be
\label{einsteinp}
\ka^{2p}\,G_{(p)}{}_{\mu}^m=\frac{(4p)!}{(2p)((2p)!)^2}\,\Lambda\,e_{\mu}^m\,.
\ee
The consistency condition of the double selfduality condition \re{dsd} and the
field equation \re{einsteinp} is
\be
\label{consistencyp}
\ka^{2p}\,R_{(p)}\ =\ -2\,\frac{(4p)!}{((2p)!)^2}\,\Lambda\,,
\ee
implying that if the $p$-th order Ricci scalar is in this way related to the
cosmological constant $\Lambda$, then such a solution of the double selfduality
equation satisfies also the Einstein equation. This is independent of the sign
of $\Lambda$, and is of course true also for the particular case of vanishing
cosmological constant $\Lambda=0$.

Next we consider a spacetime of dimension $d=2(p+q)$, with $q>p$.
The double selfduality condition in this case is
\be
\label{dsdpq}
R_{\mu_1\mu_2....\mu_{2p}}^{m_1m_2..m_{2p}}=\pm\ka^{2(q-p)}
\frac{e}{[(2q)!]^2}\,
\vep_{\mu_1\mu_2...\mu_{2p}\nu_1\nu_2...\nu_{2q}}\,
R^{\nu_1\nu_2....\nu_{2p}}_{n_1n_2..n_{2q}}\,
~\vep^{m_1m_2...m_{2p}n_1n_2...n_{2q}}\,,
\ee
where $\ka$ is a constant with dimensions of a length.

Let us contract the l.h.s. of \re{dsdpq} with
$e^{\mu_2}_{m_2}\,e^{\mu_2}_{m_2}\,...\,e^{\mu_{2p}}_{m_{2p}}$, and let us
relabel the free indices $(\mu_1,m_1)$ as $(\mu,m)$. This yields the $p$-th
order Ricci tensor defined by \re{ricci-t}, which after applying the usual
tensor identities results in the constraint
\be
\label{dsd-pq}
R_{(p)}{}_{\mu}^m\ =\ \mp\ka^{2(q-p)}\frac{(2p-1)!}{(2q-1)!}\ G_{(q)}{}_{\mu}^m\,.
\ee
Before comparing this constraint with the Einstein equations of the appropriate
gravitational system (a subsystem of \re{EHhier} plus a cosmological constant), it
is convenient to state the corresponding constraint arising from the inverse of the
double selfduality constraint \re{dsdpq}, namely of
\be
\label{dsdqp}
R_{\mu_1\mu_2....\mu_{2q}}^{m_1m_2..m_{2q}}=\pm\ka^{2(p-q)}
\frac{e}{[(2p)!]^2}
\vep_{\mu_1\mu_2...\mu_{2q}\nu_1\nu_2...\nu_{2p}}
R^{\nu_1\nu_2....\nu_{2q}}_{n_1n_2..n_{2p}}
~\vep^{m_1m_2...m_{2q}n_1n_2...n_{2p}}\,.
\ee
This is
\be
\label{dsd-qp}
R_{(q)}{}_{\mu}^m=\mp\ka^{2(p-q)}\frac{(2q-1)!}{(2p-1)!}\ G_{(p)}{}_{\mu}^m\,.
\ee
Again we consider the case of Minkowskian signature (the lower sign) first, to
dispose of it as above. In this case \re{dsd-pq} and \re{dsd-qp} simply yield
\be
\label{weakpq}
(2q)!\,\ka^{2p}\,R_{(p)}\ +\ (2p)!\,\ka^{2q}\,R_{(q)}\ =\ 0\,,
\ee
which is too weak to solve an Einstein equations. Hence again we restrict to the
Euclidean signature case.

Now the gravitational system in Euclidean space appropriate to \re{dsd-pq} and
\re{dsd-qp} is
\be
\label{EHpq}
{\cal L}_{\rm{grav}}\ =
\ e\ \left(\frac12\ka_1^{2p} \ R_{(p,q)}\ +\ \frac12\ka_2^{2q} \ R_{(q,p)}
\ +\ d!\ \Lambda\right)\,,
\ee
whose Einstein equations are
\be
\label{pqeinstein}
\ka_1^{2p}\ (2p)\ G_{(p)}{}_{\mu}^m\ +\ \ka_2^{2q}\ (2q)\ G_{(q)}{}_{\mu}^m\ =\
\frac{(2(p+q))!}{(2p)!\,(2q)!}\ \Lambda\ e_{\mu}^m\,.
\ee
The consistency conditions arising from the identification of \re{dsd-pq} with
\re{pqeinstein}, and \re{dsd-qp} with \re{pqeinstein}, are
\be
\label{consistencypq}
\ka_1^{2p}\,R_{(p)}\,=\,\ka_2^{2q}\,R_{(q)}\,
=\,-2\,\frac{(2(p+q))!}{(2p)!\,(2q)!}\ \Lambda
\ee
and
\be
\label{kappas}
\ka^{2(q-p)}=\frac{(2q)!}{(2p)!}\,\frac{\ka_2^{2q}}{\ka_1^{2p}}\,.
\ee
Condition \re{consistencypq} is the analogue of \re{consistencyp}, to which it
reduces when one sets $p=q$, while
\re{kappas} is an additional condition in this case constraining the relative values
of the three constants $\ka_1$, $\ka_2$ and $\ka$, with dimensions of length.

Again, provided \re{kappas} is satisfied, the two conditions \re{consistencypq}
imply that the Einstein equation is satisfied by the solutions of the double
selfduality equations.
This completes our discussion of double-selfdual spaces.

%%%%%%%%%%%%%%%%%%%%%%%%%%%%%%%%%%%%%%%%%%%%%%%%%%%%%%%%%%%%%%%%%%%%%%%%%%%%%%%
%\begin{small}

\end{document}